\documentclass[a4paper,12pt]{article}

\usepackage{amsmath}

\begin{document}

\begin{titlepage}
\begin{flushright}
EPHOU-03-006 \\
October, 2003
\end{flushright}

\vspace{5cm}

\begin{center}
{\Large Twisted Superspace for N=D=2 Super BF and Yang-Mills 
}\\
\vspace{0.3cm}
{\Large with Dirac-K\"ahler Fermion Mechanism}

\vspace{1cm}

{\sc Junji Kato}\footnote{jkato@particle.sci.hokudai.ac.jp}, 
{\sc Noboru Kawamoto}\footnote{kawamoto@particle.sci.hokudai.ac.jp} and 
{\sc Yukiya Uchida}\\

{\it{ Department of Physics, Hokkaido University }}\\
{\it{ Sapporo, 060-0810, Japan}}\\
\end{center}

\vspace{2cm}

\begin{abstract}
We propose a twisted D=N=2 superspace formalism. 
The relation between the twisted super charges including 
the BRST charge, vector and pseudo scalar super charges 
and the N=2 spinor super charges is established. 
We claim that this relation is essentially related with the 
Dirac-K\"ahler fermion mechanism. 
We show that a fermionic bilinear form of twisted N=2 chiral 
and anti-chiral superfields 
is equivalent to the quantized version of BF theory with the Landau 
type gauge fixing while a bosonic bilinear form leads to the 
N=2 Wess-Zumino action. 
We then construct a Yang-Mills action 
described by the twisted N=2 chiral and vector superfields, and show that 
the action is equivalent to the twisted version of the D=N=2 super Yang-Mills 
action, previously obtained from the quantized generalized topological 
Yang-Mills action with instanton gauge fixing.
\end{abstract}

\end{titlepage}

\newpage

\setcounter{equation}{0}
\renewcommand{\theequation}{\arabic {section}.\arabic{equation}}

\section{Introduction}

It is obviously one of the most fundamental issues to understand the origin 
of super symmetry if any. The topological field theory proposed by 
Witten\cite{W} posed a possibility that N=2 Yang-Mills action can be 
generated by the twisted version of super Yang-Mills action which is 
equivalent to the quantized topological Yang-Mills theory. 
It was soon recognized that Witten's formulation could be derived from 
the ``partially'' BRST gauge fixed action of the topological Yang-Mills 
action with instanton gauge-fixing~[2-5].
Here the twisting procedure played an important role to generate 
matter fermions from the ghost related fermions.
After this proposal there have been many systematic 
investigations on this subject[6-12],
discovery of a new type of supersymmetry[13-27],
related topological actions[28-32]
and a new type of topological actions\cite{KW,KW2}. 

In the investigations of the topological field theories of Schwarz type; 
Chern-Simons action and BF actions, a new type of vector supersymmetry was 
discovered[13-16].
It was recognized that this vector 
supersymmetry belongs to a twisted version of an extended supersymmetry of 
N=2 or N=4. The origin of the vector supersymmetry was recognized 
in some particular examples to be related to the fact that the energy-momentum 
tensor can be expressed as a pure BRST variation[35-37].
It was later stressed that a (pseudo) scalar supersymmetry 
was also accompanied together with BRST and vector supersymmetry\cite{CLS}. 
The connection of the extended supersymmetry and the quantization procedure of 
anti-field formalism by Batalin and Vilkovisky was also 
investigated\cite{DR,BDL}. 

Even with those intensive investigations on the relation between the quantized 
topological field theories and twisted supersymmetry, there still remained 
unclear the fundamental understandings of the relations between the 
BRST symmetry and the vector and (pseudo) scalar supersymmetry, 
quantization of topological field theories, and the twisting mechanism. 

In the previous paper\cite{KT} one of the authors and Tsukioka showed that 
the two dimensional twisted N=2 super Yang-Mills action can be derived 
by instanton gauge fixing of the two dimensional version of the topological 
Yang-Mills action of the generalized gauge theory. 
This quantized action has the close connection with N=2 super Yang-Mills 
action obtained from the different context\cite{DVF,SchapT}.
This two dimensional 
formulation is completely parallel to the Witten's four dimensional 
version of the topological field theory.  
In the paper it was explicitly shown how the BRST charge, the charges of 
the vector supersymmetry and the pseudo scalar supersymmetry  
constitute twisted N=2 super charges and are related to 
the spinor super charges of N=2 supersymmetry in two dimensions.  
The twisting procedure generating the 
matter fermions from ghost related fields is essentially equivalent 
to the Dirac-K\"ahler fermion formulation\cite{KT}. The R-symmetry 
of N=2 super symmetry is nothing but the "flavor" symmetry of the 
Dirac-K\"ahler fermion fields. 

The mysterious relations between the super charges of N=2 supersymmetry 
and BRST charges and the newly discovered vector and peudo schalar 
supersymmetry charges are cleared up. They have the same relations 
as the Dirac-K\"ahler fermion fields and anti-symmetric tensor fields 
of differential forms in the Dirac-K\"ahler fermion formulation. 
The mechanism how the matter fermions are generated from the ghost related 
fields by the twisting procedure is essentially Dirac-K\"ahler fermion 
mechanism itself, {\it i.e.} the matter fermions are generated by the 
ghost related fields which have anti-symmetric tensor suffixes.

In this article we show that there exists twisted superspace formalism 
which naturally accommodates the twisting procedure and the Dirac-K\"ahler 
fermion mechanism. So far most of the examples of the twisted supersymmetric 
models have only on-shell N=2 or N=4 supersymmetry. We show that 
the quantized BF and super Yang-Mills actions can have off-shell N=2 
supersymmetry by introducing auxiliary fields. We show that those 
quantized BF and Yang-Mills actions can be equivalently formulated by 
a simple form of twisted N=2 chiral super fields and vector superfield. 
We can thus establish the twisted superspace formalism in two dimensions 
which we claim as the most essential formulation behind the twisted 
supersymmetry.    

In this article Dirac-K\"ahler fermion formulation plays a fundamental 
role. The original idea that fermion field can be formulated by 
differential forms is old back to Ivanenko and Landau\cite{IL}.
It was later shown by K\"ahler~\cite{Kahler} that Dirac equation is 
constructed from the direct sum of inhomogeneous differential forms 
which is called Dirac-K\"ahler field[45-50].
Here all the degrees of differential forms are needed to express the 
Dirac-K\"ahler fields with "flavor" suffix which is shown to denote 
the extended supersymmetry suffix as well. 

In formulating the two dimensional super Yang-Mills action, 
we needed to start from the two dimensional version of topological 
Yang-Mills action of the generalized gauge theory.  
The generalized gauge theory is the generalization of the standard 
gauge theory by introducing all the degrees of differential 
forms and can be defined in arbitrary dimensions. In particular the 
generalization of the Chern-Simons actions into arbitrary dimensions 
was proposed by one of the authors (N.K.) and Watabiki\cite{KW,KW2}. 
It is interesting to note that the generalized gauge theory also
introduce all the degrees of differential forms 
as gauge fields and parameters together with quaternion 
structure. 

This paper is organized as follows.  
In section 2 we show that the quantized BF models with auxiliary fields 
in two dimensions have off-shell N=2 twisted super symmetry. 
In section 3 we quantize the generalized two-dimensional 
topological Yang-Mills theory with the instanton gauge fixing 
and show that the quantized action leads to twisted N=2 super 
Yang-Mills action at the on-shell level. 
In section 4 we formulate the twisted N=2 super space formalism for chiral 
super fields and vector super fields and show that 
the quantized BF actions and the quantized topological Yang-Mills action 
of generalized gauge theory in two dimensions 
can be written down by chiral and vector superfields and thus possess 
off-shell N=2 twisted super symmetry. In section 5 we explain that the 
twisting mechanism is essentially equivalent to the Dirac-K\"ahler fermion 
formulation.  
Conclusions and discussions are given in the final section.


\section{A simple model of twisted N=D=2}
\setcounter{equation}{0}

\subsection{Twisted D=N=2 Supersymmetry Algebra}
In this subsection, we summarize the construction of the twisted D=N=2
supersymmetry algebra\cite{bms,lp}. Throughout this paper, we consider the 
two-dimensional Euclidean space-time. The notation is summarized in 
Appendix. 

We first introduce D=N=2 supersymmetry algebra without a central extension:
\begin{equation}
\begin{split}
\{Q_{\alpha i},Q_{\beta j} \}
 &=2\delta_{ij} {\gamma^\mu}_{\alpha \beta}P_\mu,\\
[P_\mu,Q_{\alpha i}]&=0,\\
[J,Q_{\alpha i}]&=\tfrac{i}{2}(\gamma^5)_\alpha{^\beta}Q_{\beta i},\\
[R,Q_{\alpha i}]&=\tfrac{i}{2}(\gamma^5)_i{^j}Q_{\alpha j},\\
[J,P_\mu]&=i\epsilon_\mu{^\nu}P_\nu,\\
[P_\mu,P_\nu]&=[P_\mu,R]=[J,R]=0.
\end{split}
\label{eq:N=2 algebra}
\end{equation}
Here $Q_{\alpha i}$ is super-charge, where the left-indices $\alpha(=1,2)$ 
and the
right-indices $i(=1,2)$ are Lorentz spinor and internal spinor suffixes 
labeling two different N=2 super-charges, respectively. We can take 
these operators 
to be Majorana. $P_\mu$ is generator of translation. $J$ and $R$ are
generators of $SO(2)$ Lorentz and $SO(2)_I$ internal rotation called $R$
symmetry, respectively.

The essential meaning of the topological twist is to identify the isospinor
indices as the spinor ones. Then isospinor should transform as spinor under
the Lorentz transformation. This will lead to a redefinition of the
energy-momentum tensor and the Lorentz rotation generator.

We can redefine the energy-momentum tensor $T_{\mu\nu}$ as the following
relation without breaking the conservation law:
\begin{align}
T^\prime_{\mu\nu}
=T_{\mu\nu}+\epsilon_{\mu\rho}\partial^\rho R_\nu
+\epsilon_{\nu\rho}\partial^\rho R_\mu,
\end{align}
where $R_\mu$ is the conserved current associated with 
$R$ symmetry\cite{W,EY,LL}. This
modification leads to a redefinition of the Lorentz generator,
\begin{align}
J^\prime=J+R.
\end{align}
This rotation group is interpreted as the diagonal subgroup of $SO(2)\times
SO(2)_I$.

Now the super charges have double spinor indices and thus can be decomposed
into the following scalar, vector and pseudo-scalar components:
\begin{align}
{\bf Q}_{\alpha \beta}= \left(\mathbf{1} Q + \gamma^\mu Q_\mu + \gamma^5
\tilde{Q}\right)_{\alpha\beta},
\end{align}
or equivalently,
\begin{equation}
\begin{split}
Q
  &=\frac{1}{2}\hbox{Tr} ({\bf Q}) \\
Q_\mu
  &=\frac{1}{2}\hbox{Tr} (\gamma_\mu {\bf Q}) \\
\tilde{Q}&
 =-\frac{1}{2}\hbox{Tr} (\gamma^5 {\bf Q})
\end{split}
\end{equation}

The relations (\ref{eq:N=2 algebra}) can be rewritten by the twisted
generators:
\begin{equation}
\begin{split}
\{Q,Q_\mu\}&=P_\mu,\
\{\tilde{Q},Q_\mu\}=-\epsilon_{\mu\nu}P^\nu,\
Q^2=\tilde{Q}^2=\{Q,\tilde{Q}\}=\{Q_\mu,Q_\nu\}=0,\\
[Q,P_\mu]&=[\tilde{Q},P_\mu]=0,\ [Q_\mu,P_\nu]=0,\\
[J^\prime,Q]&=[J^\prime,\tilde{Q}]=0,\
[J^\prime,Q_\mu]=i\epsilon_{\mu\nu}Q^\nu,\\
[R,Q]&=\tfrac{i}{2}\tilde{Q},\
[R,Q_\mu]=\tfrac{i}{2}\epsilon_{\mu\nu}Q^\nu,\
[R,\tilde{Q}]=-\tfrac{i}{2}Q,\ \\
[J^\prime,P_\mu]&=i\epsilon_{\mu\nu}P^\nu,\ \\
[P_\mu,P_\nu]&=[P_\mu,R]=[J^\prime,R]=0.
\end{split}
\label{eq:N=2 T-algebra}
\end{equation}
This is the twisted D=N=2 supersymmetry algebra.


\subsection{N=D=2 Super BF from quantized BF theory}

In order to reveal the fundamental relation between the quantization of 
topological action and supersymmetry, we consider the simplest example 
of the two-dimensional abelian BF theory\cite{BBRT,BT}. 
The action is given by 
\begin{align}
S_{ABF}
=\int_{M_2} d^2x 
 \epsilon^{\mu\nu}\phi\partial_\mu\omega_\nu
\end{align}
where $M_2$ is a two-dimensional Euclidean manifold,
$\epsilon^{12}=\epsilon_{12}=1$, and $\phi$ is a $0$-form field. This action
has the following obvious gauge invariance:
\begin{align}
\begin{split}
\delta \phi &=0,\\
\delta \omega_\mu &=\partial_\mu v.
\end{split}
\end{align}

To obtain the gauge fixed action, we introduce the ghost field $c$,
anti-ghost field $\bar{c}$, and the auxiliary field $b$, and define their
BRST transformation as follows:
\begin{align}
\begin{split}
s \phi &=0,\\
s \omega_\mu &=\partial_\mu c,\\
s c &= 0,\\
s \bar{c} &= -ib,\\
s b &=0,
\end{split}
\label{eq:BF BRST}
\end{align}
where $s^2=0$. Then we can obtain the gauge fixed action in the Landau
gauge,
\begin{align}
\begin{split}
S_{\hbox{on-shell AQBF}}&=\int_{M_2} d^2x 
 [\epsilon^{\mu\nu}\phi\partial_\mu\omega_\nu
 +is(\bar{c}\partial^\mu\omega_\mu)],\\
&=\int_{M_2} d^2x 
 [\epsilon^{\mu\nu}\phi\partial_\mu\omega_\nu
 +b\partial^\mu\omega_\mu
 -i\bar{c}\partial^\mu \partial_\mu c].
\end{split}
\label{eq:QBF action}
\end{align}
This action has not only the BRST symmetry (\ref{eq:BF
BRST}), but also has two more fermionic symmetries[13-27] 
as shown in the following table 1.
\begin{table}[htbp]
\begin{center}
\begin{tabular}{c|c|c|c}
\hline
$\phi^A$ & $s\phi^A$ & $s_\mu \phi^A$ & $\widetilde{s}\phi^A$ \\
\hline
$\phi$
 & $0$
 & $-\epsilon_{\mu\nu}\partial^\nu\bar{c}$ 
 & $0$ \\
$\omega_\nu$
 & $\partial_\nu c$
 & $0$
 & $-\epsilon_{\nu\rho}\partial^\rho c$\\
$c$
 & $0$
 & $-i\omega_\mu$
 & $0$\\
$\bar{c}$
 & $-ib$
 & $0$
 & $-i\phi$\\
$b$
 & $0$
 & $\partial_\mu\bar{c}$
 & $0$\\
\hline
\end{tabular}
\caption{On-shell N=2 twisted super transformation of abelian BF model.}
\end{center}
\end{table}
We can see that these operators satisfy the following relations if the
equations of motion holds:
\begin{align}
\begin{split}
s^2&=\{s,\tilde{s}\}=\tilde{s}^2=\{s_\mu,s_\nu\}=0,\\
\{s,s_\mu\}&=-i\partial_\mu,\{\tilde{s},s_\mu\}=
i\epsilon_{\mu\nu}\partial^\nu.
\end{split}
\end{align}
This is the twisted D=N=2 supersymmetry algebra which can be recognized 
by defining the super charge operators,
\begin{align}
Q_{\alpha i}= \left(\mathbf{1} s + \gamma^\mu s_\mu + \gamma^5
\tilde{s}\right)_{\alpha i},
\label{supercharge-twist}
\end{align}
which satisfy the following extended N=2 supersymmetry:
\begin{align}
\{Q_{\alpha i},Q_{\beta j} \}
 =2\delta_{ij} {\gamma^\mu}_{\alpha\beta}P_\mu. 
\label{supercharge-algebra}
\end{align}
Thus the quantuzed BF theory fixed in Landau gauge has the twisted 
D=N=2 supersymmetry at the on-shell level.

By adding auxilialy fields to eq.(\ref{eq:QBF action}) we can find 
off-shell N=2 supersymmetric action 
\begin{align}
S_{\hbox{off-shell AQBF}}=\int_{M_2} d^2x 
 [\epsilon^{\mu\nu}\phi\partial_\mu\omega_\nu
 +b\partial^\mu\omega_\mu
 -i\bar{c}\partial^\mu \partial_\mu c -i\lambda\rho],
\label{eq:QBF action2}
\end{align}
which has the off-shell extended N=2 supersymmetry invariance shown in 
Table 2.
\begin{table}[htbp]
\begin{center}
\begin{tabular}{c|c|c|c}
\hline
$\phi^A$ & $s\phi^A$ & $s_\mu \phi^A$ & $\widetilde{s}\phi^A$ \\
\hline
$\phi$
 & $i\rho$
 & $-\epsilon_{\mu\nu}\partial^\nu\bar{c}$ 
 & $0$ \\
$\omega_\nu$
 & $\partial_\nu c$
 & $-i\epsilon_{\mu\nu}\lambda$
 & $-\epsilon_{\nu\rho}\partial^\rho c$\\
$c$
 & $0$
 & $-i\omega_\mu$
 & $0$\\
$\bar{c}$
 & $-ib$
 & $0$
 & $-i\phi$\\
$b$
 & $0$
 & $\partial_\mu\bar{c}$
 & $-i\rho$\\
$\lambda$
 & $\epsilon^{\mu\nu}\partial_\mu\omega_\nu$ 
 & $0$
 & $-\partial_\mu\omega^\mu$\\
$\rho$
 & $0$
 & $-\partial_\mu\phi-\epsilon_{\mu\nu}\partial^\nu b$
 & $0$\\
\hline
\end{tabular}
\caption{Off-shell N=2 twisted super transformation of abelian BF model.}
\end{center}
\end{table}

It turns out that the off-shell N=2 supersymmetric action is equivalent to the 
following simple form:
\begin{align}
S_{\hbox{off-shell AQBF}}=\int_{M_2} d^2x 
s\tilde{s}\frac{1}{2}\epsilon^{\mu\nu}s_\mu s_\nu (-i\bar{c}c).
\label{eq:QBF action3}
\end{align}
This form of action suggests that there is a superspace formulation of 
extended twisted N=2 supersymmetry.

The extension from abelian group to non-abelian group is straightforward. 
The off-shell N=2 supersymmetric quantized BF action with non-abelian 
gauge group is given by: 
\begin{align}
\begin{split}
S_{\hbox{off-shell NABQBF}}
=\int_{M_2} d^2x \ \hbox{Tr} \ 
 [\phi F
 +b\partial^\mu\omega_\mu
 +i\partial^\mu \bar{c} D_\mu c -i\lambda\rho],
\end{split}
\label{eq:NABQBF action1}
\end{align}
where the fields are all Lie algebra valued and
\begin{eqnarray}
F &=& \epsilon^{\mu\nu}(\partial_\mu\omega_\nu + \omega_\mu\omega_\nu)\\
D_\mu &=& \partial_\mu c + [\omega_\mu,c].
\end{eqnarray}

The off-shell extended N=2 supersymmetry transformation is given in 
Table 3.

\begin{table}[htbp]
\begin{center}
\begin{tabular}{c|c|c|c}
\hline
$\phi^A$ & $s\phi^A$ & $s_\mu \phi^A$ & $\widetilde{s}\phi^A$ \\
\hline
$\phi$
 & $[\phi,c]+i\rho$
 & $-\epsilon_{\mu\nu}\partial^\nu\bar{c}$ 
 & $0$ \\
$\omega_\nu$
 & $D_\nu c$
 & $-i\epsilon_{\mu\nu}\lambda$
 & $-\epsilon_{\nu\rho}\partial^\rho c$\\
$c$
 & $-c^2$
 & $-i\omega_\mu$
 & $0$\\
$\bar{c}$
 & $-ib$
 & $0$
 & $-i\phi$\\
$b$
 & $0$
 & $\partial_\mu\bar{c}$
 & $-[\phi,c]-i\rho$\\
$\lambda$
 & $F-\{\lambda,c\}$ 
 & $0$
 & $-\partial_\mu\omega^\mu$\\
$\rho$
 & $-\{\rho,c\}$
 & $-D_\mu\phi-\epsilon_{\mu\nu}\partial^\nu b-i
 \epsilon_{\mu\nu}\{\partial^\nu\bar{c},c\}$
 & $0$\\
\hline
\end{tabular}
\caption{Off-shell N=2 twisted super transformation of non-abelian BF model.}
\end{center}
\end{table}
Just like in the abelian case this off-shell N=2 super symmetric action 
can be equivalently written down by the same simple form as the abelian 
case
\begin{align}
S_{\hbox{off-shell NABQBF}}=\int_{M_2} d^2x 
s\tilde{s}\frac{1}{2}\epsilon^{\mu\nu}s_\mu s_\nu \hbox{Tr}(-i\bar{c}c),
\label{eq:NAQBF action2}
\end{align}
which again suggests the N=2 twisted superspace formalism.


\setcounter{equation}{0}
\section{N=D=2 super Yang-Mills from quantized generalized topological 
Yang-Mills}
In this subsection, we briefly summarize the formulation of the generalized 
gauge theory introduced by Kawamoto and Watabiki\cite{KW,KW2}, and the 
twisted D=N=2 generalized topological Yang-Mills model\cite{KT}.

\subsection{Generalized gauge theory}
The generalized gauge field $\mathcal{A}$ and the generalized gauge
parameter $\mathcal{V}$ are defined as follows:
\begin{align}
\mathcal{A}
&= \mathbf{1}\psi+\mathbf{i}\hat{\psi}+\mathbf{j}A+\mathbf{k}\hat{A}, \\
\mathcal{V}
&= \mathbf{1}\hat{a}+\mathbf{i}a+\mathbf{j}\hat{\alpha}+\mathbf{k}\alpha.
\end{align}
Here $(\psi,\alpha)$, $(\hat{\psi},\hat{\alpha})$, $(A,a)$, and
$(\hat{A},\hat{a})$ are the direct sums of fermionic odd-forms, fermionic
even-forms, bosonic odd-forms, and bosonic even-forms, respectively, and
they are defined as the following graded Lie algebra valued fields:
\begin{align}
A =A^a T_a ,\
\hat{A} =\hat{A}^\alpha\Sigma_\alpha ,\
\psi =\psi^\alpha \Sigma_\alpha ,\
\hat{\psi} = \hat{\psi}^a T_a ,\\
\hat{a} = \hat{a}^a T_a ,\
a = a^\alpha\Sigma_\alpha ,\
\hat{\alpha} = \hat{\alpha}^\alpha \Sigma_\alpha ,\
\alpha =\alpha^a T_a,
\end{align}
where the generators $T_a$ and $\Sigma_\alpha$ satisfy the relations
\begin{align}
[T_a,T_b]=f_{ab}^cT_c\ ,\
[T_a,\Sigma_\beta]=g_{a\beta}^\gamma\Sigma_\gamma\ ,\
\{\Sigma_\alpha,\Sigma_\beta\}=h_{\alpha\beta}^cT_c.
\end{align}
The symbols $\mathbf{1}$, $\mathbf{i}$, $\mathbf{j}$, and $\mathbf{k}$
satisfy the following quaternion algebra:
\begin{align}
\mathbf{1}^2&=\mathbf{1} ,
\mathbf{i}^2=\mathbf{j}^2=\mathbf{k}^2=-\mathbf{1} , \\
\mathbf{i}\mathbf{j}&=-\mathbf{j}\mathbf{i}=\mathbf{k} ,
\mathbf{j}\mathbf{k}=-\mathbf{k}\mathbf{j}=\mathbf{i} ,
\mathbf{k}\mathbf{i}=-\mathbf{i}\mathbf{k}=\mathbf{j}.
\end{align}

The generalized Chern-Simons action on the even and odd dimensional manifold
are given
\begin{align}
S_{even}
 &=\int_{M_{even}} \text{Tr}_{\mathbf{k}}\left(
 \mathcal{A}\mathcal{Q}\mathcal{A}
 +\frac{2}{3}\mathcal{A}^3
 \right),
\label{eq:even CS}\\
S_{odd}
 &=\int_{M_{odd}} \text{Str}_{\mathbf{j}}\left(
 \mathcal{A}\mathcal{Q}\mathcal{A}
 +\frac{2}{3}\mathcal{A}^3
 \right),
\label{eq:odd CS}
\end{align}
where the traces Tr$_{\mathbf{k}}(\cdots)$ and Str$_{\mathbf{k}}(\cdots)$
are defined so as to pick up the coefficient of $\mathbf{k}$ and
$\mathbf{j}$ from $(\cdots)$, respectively, and fulfill the following
characteristic of traces:
\begin{align}
\begin{split}
\text{Tr}[T_a,\cdots]&=\text{Tr}[\Sigma_\alpha,\cdots]=0,\\
\text{Str}[T_a,\cdots]&=\text{Str}\{\Sigma_\alpha,\cdots\}=0,
\end{split}
\end{align}
where the number of $\Sigma_\alpha$s in $\cdots$ of
$\{\Sigma_\alpha,\cdots\}$ should be odd. The differential operator
$\mathcal{Q}$ is defined as follows:
\begin{align}
\mathcal{Q}=\mathbf{j}d.
\end{align}
where $d=dx^\mu\partial_\mu$.

The generalized Chern-Simons actions (\ref{eq:even CS}) and (\ref{eq:odd
CS}) are invariant under the following gauge transformation:
\begin{equation}
\delta \mathcal{A}=[\mathcal{Q}+\mathcal{A},\mathcal{V}].
\end{equation}
This symmetry is larger than the usual gauge symmetry since the generalized
gauge parameter $\mathcal{V}$ contains the various form parameters.

As in the usual gauge theory, it is possible to define the generalized Chern
character:
\begin{align}
\text{Str}_{\mathbf{1}}(\mathcal{F}^n)
 &=\text{Str}_{\mathbf{1}}(\mathcal{Q}\Omega_{2n-1}),
 \label{eq:even F}\\
\text{Tr}_{\mathbf{i}}(\mathcal{F}^n)
 &=\text{Tr}_{\mathbf{i}}(\mathcal{Q}\Omega_{2n-1}),
 \label{eq:odd F}
\end{align}
where $\mathcal{F}$ is the generalized curvature
\begin{align}
\mathcal{F}=\mathcal{Q}\mathcal{A}+\mathcal{A}^2,
\end{align}
and $\Omega_{2n-1}$ is the generalized Chern-Simons form. Equations
(\ref{eq:even F}) and (\ref{eq:odd F}) are bosonic even form and bosonic odd
form, respectively. In case of $n=2$, the topological Yang-Mills type action
can be obtained from the generalized Chern-Simons Lagrangian in an
even-dimensional manifold $M$ as follows:
\begin{align}
S_{TYM}
=\int_M \text{Str}_{\mathbf{1}}(\mathcal{F}^2)
=\int_M \text{Str}_{\mathbf{1}}\left[
 \mathcal{Q}\left(
  \mathcal{A}\mathcal{Q}\mathcal{A}+\frac{2}{3}\mathcal{A}^3
 \right)\right].
\end{align}


\subsection{Quantization of generalized topological Yang-Mills action in D=2}

The two dimensional generalized gauge field without fermionic components can
be taken as the form
\begin{align}
\mathcal{A}_0=\mathbf{j}\omega^aT^a+\mathbf{k}
(\phi^\alpha+B^\alpha)\Sigma^\alpha,
\end{align}
where $\phi$, $\omega$, and $B$ are graded Lie algebra valued bosonic 0-,
1-, and 2-form field, respectively. We can take the following algebra as the
graded Lie algebra:
\begin{align}
\{T^a\}=\{1,\gamma^5\}, \{\Sigma^\alpha\}=\{\gamma^1,\gamma^2\},
\end{align}
then $\text{Str}$ can be taken
\begin{align}
\text{Str}(\cdots)=\text{Tr}(\gamma^5\cdots).
\end{align}
The generalized curvature is given by
\begin{equation}
\begin{split}
\mathcal{F}_0
 &=\mathcal{Q}\mathcal{A}_0+{\mathcal{A}_0}^2\\
 &=-\mathbf{1}(d\omega+\phi^2+\{\phi,B\})+\mathbf{i}(d\phi+[\omega,\phi]).
\end{split}
\end{equation}

Now two-dimensional topological Yang-Mills action is given as follows:
\begin{align}
\begin{split}
S_{TYM}
&= \frac{1}{2} \int \text{Str}_{\mathbf{1}} {\mathcal{F}_0}^2 
=\frac{1}{2} \int \text{Str}_{\mathbf{1}}\left[
 \mathcal{Q}\left(
  \mathcal{A}_0\mathcal{Q}\mathcal{A}_0+\frac{2}{3}\mathcal{A}_0^3
 \right)\right] \\
&= \int d^2x \epsilon^{\mu\nu}(F_{\mu\nu}|\phi|^2
 +\epsilon^{ab}(D_\mu\phi)_a(D_\nu\phi)_b)\\
&= \int d^2x \epsilon^{\mu\nu}\partial_\mu (
 2\omega_\nu|\phi|^2+\epsilon^{ab}\phi_a\partial_\nu\phi_b),
\end{split}
\label{eq:TYM action}
\end{align}
where $F_{\mu\nu}=\partial_\mu\omega_\nu-\partial_\nu\omega_\mu$,
$|\phi|^2=\phi^a\phi_a$, and
$(D_\mu\phi)_a=\partial_\mu\phi_a-2\epsilon_{ab}\omega_\mu\phi^b$. In this
action, the scalar component of even form generators for the 1-form field
and odd form generator components of the 2-form field drop out due to the
reducible structures of the gauge transformations. Then the generalized
gauge transformation turns out to be the following $SO(2)$ gauge
transformation:
\begin{align}
\begin{split}
\delta_{gauge} \phi_a &= 2v\epsilon_{ab}\phi^b,\\
\delta_{gauge} \omega_\mu &= \partial_\mu v,
\end{split}
\end{align}
where $v$ is the 0-form gauge parameter.

The Topological Yang-Mills action (\ref{eq:TYM action}) has the topological
shift symmetry:
\begin{equation}
\begin{split}
\delta_{shift} \phi_a &= u_a,\\
\delta_{shift} \omega_\mu &= u_\mu,
\end{split}
\end{equation}
where $u_a$ and $u_\mu$ are the shift parameters.

We can see that the action (\ref{eq:TYM action}) is invariant under the
following BRST transformation:
\begin{align}
\begin{split}
s \phi_a &= 2\epsilon_{ab}\phi^bC-\tilde{C}_a,\\
s \omega_\mu &=\partial_\mu C+\tilde{C}_\mu,\\
s C &=-\eta,\\
s \tilde{C}_a &=2\epsilon_{ab}(C\tilde{C}^b-\phi^b \eta),\\
s \tilde{C}_\mu &= \partial_\mu \eta,\\
s \eta &=0,
\end{split}
\end{align}
where $C$, $(\tilde{C}_a,\tilde{C}_\mu)$, and $\eta$ are the ghost
associated with the $SO(2)$ gauge parameter, and the ghosts of the
topological shift symmetry, and the ghost for the ghost of the reducible
gauge symmetry, respectively.

We can find a two-dimensional instanton relation of the generalized gauge
system by imposing the self- (anti-self-) dual condition:
\begin{align}
*\mathcal{F}_0=\pm\mathcal{F}_0.
\end{align}
Here the dual operator $*$ operates on the differential forms as the Hodge
dual operation, furthermore the dual of the generators and the quaternions
are defined as follows:
\begin{align}
*1&=-\gamma^5,\ *\gamma^a=\epsilon^{ab}\gamma_b,\ *\gamma^5=-1,\\
*\mathbf{1}&=\mathbf{1},\ *\mathbf{i}=-\mathbf{i}.
\end{align}

Then we can find the following minimal condition of the action leading to
the instanton relations:
\begin{align}
\begin{split}
&\frac{1}{2} \int \text{Str}_{\mathbf{1}}
 \left(
  \pm \mathcal{F}_0\wedge\mathcal{F}_0
  + \mathcal{F}_0\wedge * \mathcal{F}_0
 \right)\\
&=\int d^2x
 \left(
  (F\pm |\phi|^2)^2
  +2(D_\mu\phi)_a^{(\pm)}(D^\mu\phi)^{(\pm)a}
 \right),
\end{split}
\end{align}
where $F=\frac{1}{2}\epsilon_{\mu\nu}F^{\mu\nu}$, and
$(D_\mu\phi)_a^{(\pm)}=\frac{1}{2}[(D_\mu\phi)_a\pm\epsilon_{\mu\nu}\epsilon
_{ab}(D^\nu\phi)^b]$. The instanton relations are obtained  from the
conditions for the absolute minima of the generalized Yang-Mills action:
\begin{align}
F- |\phi|^2=0,\ (D_\mu\phi)_a^{(-)}=0.
\label{eq:instanton}
\end{align}

Now we consider the following ``partially'' gauge fixed action:
\begin{equation}
\begin{split}
S=S_0 &+s \int d^2x \left( \bar{\eta}\partial^\mu \tilde{C}_\mu \right)\\
&+ s \int d^2x \biggl[
 \lambda\left(
   F-|\phi|^2-\frac{1}{4}\tilde{\pi}
 \right)
 -\chi^{\mu a}\left(
  (D_\mu\phi)_a^{(-)}+\frac{1}{8}\pi_{\mu a}
 \right) \biggr]\\
& +s \int d^2x \left( -2i\epsilon^{ab}\bar{\eta} \phi_a \tilde{C}_b \right)
\end{split}
\label{eq:SYM action}
\end{equation}
where
\begin{equation}
\begin{split}
s \bar{\eta} = \rho,\quad & \quad s \rho =0, \\
s \lambda = \tilde{\pi},\quad & \quad s\tilde{\pi}=0,\\
s \chi_{\mu a} = \pi_{\mu a},\quad & \quad s \pi_{\mu a}=0.
\end{split}
\end{equation}
Here $(\bar{\eta},\lambda,\chi_{\mu a})$ and $(\rho,\tilde{\pi},\pi_{\mu
a})$ are the anti-ghost fields and auxiliary fields, respectively. And the
second, and the third terms in the right-hand-side of (\ref{eq:SYM action})
correspond to the condition of the gauge fixing
$\partial^\mu\tilde{C}_\mu=0$ (the term ``partially'' means that the degree
of freedom for the $SO(2)$ gauge parameter $v$ is unfixed), and the
instanton relations (\ref{eq:instanton}), respectively. The anti-self-dual
fields $\chi_{\mu a}$ and $\pi_{\mu a}$ obey the following
conditions:\begin{align}
\chi_{\mu a}=-\epsilon_{\mu\nu}\epsilon_{ab}\chi^{\nu b},\
\pi_{\mu a}=-\epsilon_{\mu\nu}\epsilon_{ab}\pi^{\nu b}.
\end{align}

After integrating $\tilde{\pi}$ and $\pi_{\mu a}$ in (\ref{eq:SYM action}),
we obtain the following action:
\begin{align}
\begin{split}
S=\int d^2x \biggl[
 & \frac{1}{2}F^{\mu\nu}F_{\mu\nu}+(D_\mu\phi)_a(D^\mu\phi)^a+|\phi|^4\\
 & +i\rho\partial_\mu\tilde{C}^\mu
  -i\lambda\epsilon^{\mu\nu}\partial_\mu\tilde{C}_\nu\\
 & -i\chi_{\mu a}(D^\mu \tilde{C})^a
  +\partial_\mu\bar{\eta}\partial^\mu\eta\\
 & -2i\epsilon^{ab}\rho \phi_a \tilde{C}_b
  -2i\lambda \phi_a \tilde{C}^a
  -2i\chi_{\mu a}\epsilon^{ab} \tilde{C}^\mu\phi_b\\
 & -\frac{i}{4}\epsilon^{\mu\nu}\chi_{\mu a}{\chi_\nu}{^a}\eta
  +2i\bar{\eta}\epsilon_{ab}\tilde{C}^a\tilde{C}^b
  +4\eta\bar{\eta}|\phi|^2
\biggr].
\end{split}
\label{eq:KT action}
\end{align}
This action is invariant under the $SO(2)$ gauge transformation:
\begin{equation}
\begin{split}
\delta_{gauge}(\phi_a,\tilde{C}_a,\chi_{\mu a})
 &=2v\epsilon_{ab}(\phi^b,\tilde{C}^b,{\chi_\mu}^b),\\
\delta_{gauge}\omega_\mu&=\partial_\mu v,\\
\delta_{gauge}(\tilde{C}_\mu,\rho,\lambda,\eta,\overline{\eta})&=0,
\end{split}
\label{eq:SO(2) trans}
\end{equation}
where $v$ is the gauge parameter.

Furthermore the action (\ref{eq:KT action}) has the fermionic symmetries
$s$, $s_\mu$, and $\tilde{s}$ defined in Table 4.
\begin{table}[htbp]
\begin{center}
\begin{tabular}{c|c|c|c}
\hline
$\phi^A$ & $s\phi^A$ & $s_\mu \phi^A$ & $\widetilde{s}\phi^A$ \\
\hline
$\phi_a$
 & $-\tilde{C}_a$ & $\tfrac{1}{2}\chi_{\mu a}$ &
$-\epsilon_{ab}\tilde{C}^b$\\
$\chi_{\nu a}$
 & $4i(D_\nu\phi)_a^{(-)}$
 & $-4i(\delta_{\mu\nu}\epsilon_{ab}
  -\epsilon_{\mu\nu}\delta_{ab})\overline{\eta}\phi^b$
 & $-4i\epsilon_{\nu\rho}(D^\rho\phi)_a^{(-)}$ \\
$\omega_\nu$
 & $\tilde{C}_\nu$
 & $-\tfrac{1}{2}(\epsilon_{\mu\nu}\lambda+\delta_{\mu\nu}\rho)$
 & $\epsilon_{\nu\rho}\tilde{C}^\rho$ \\
$\lambda$
 & $-2i(F-|\phi|^2)$
 & $2i\epsilon_{\mu\rho}\partial^\rho\overline{\eta}$
 & $0$ \\
$\tilde{C}_a$
 & $-2i\epsilon_{ab}\phi^b\eta$
 & $-2i(D_\mu\phi)_a^{(+)}$
 & $-2i\phi_a\eta$\\
$\overline{\eta}$
 & $\rho$
 & $0$
 & $-\lambda$\\
$\tilde{C}_\nu$
 & $i\partial_\nu\eta$
 & $i(F_{\mu\nu}+\epsilon_{\mu\nu}|\phi|^2)$
 & $-i\epsilon_{\nu\rho}\partial^\rho\eta$\\
$\rho$
 & $0$
 & $2i\partial_\mu\overline{\eta}$
 & $-2i(F-|\phi|^2)$\\
$\eta$
 & $0$
 & $2\tilde{C}_\mu$
 & $0$\\
\hline
\end{tabular}
\caption{On-shell N=2 twisted super transformation.}
\end{center}
\end{table}
These transformation laws construct the on-shell twisted N=2 supersymmetry
algebra: the operators $s$, $s_\mu$, and $\tilde{s}$ obey the following
relations if the equations of motion obtained from the action (\ref{eq:KT
action}) holds:
\begin{align}
\begin{split}
s^2&=i\delta_{gauge\ \eta},\\
\{s,s_\mu\}
 &=2i\partial_\mu-2i\delta_{gauge\ \omega_\mu}, \\
\{\widetilde{s},s_\mu\}
 &=-2i\epsilon_{\mu\nu}\partial^\nu+2i\delta_{gauge\
\epsilon_{\mu\nu}\omega^\nu},\\
{\widetilde{s}}^2&=i\delta_{gauge\ \eta},\\
\{s,\widetilde{s}\}&=0, \\
\{s_\mu,s_\nu\}&=-2i\delta_{\mu\nu} \delta_{gauge\ \overline{\eta}},
\end{split}
\end{align}
where $\delta_{gauge\ {\phi^A}}$ is given in (\ref{eq:SO(2) trans}).
 
The ghost field sets $(\rho,\tilde{C}_\mu,\lambda)$ and
$(\tilde{C}_a,\chi_{\mu a})$ can construct the Dirac-K\"ahler fermions as
follows:
\begin{align}
\psi &= \tfrac{1}{2}(\rho+\gamma^\mu\tilde{C}_\mu-\gamma^5\lambda), 
\label{DKtr01}\\
\chi &= \tfrac{1}{2}(-\tilde{C}_{1}+\gamma^\mu\chi_{\mu
1}-\gamma^5\tilde{C}_{2}). \label{DKtr02}
\end{align}
Indeed the kinetic terms of these fields in (\ref{eq:KT action}) can be
expressed as
\begin{align}
\begin{split}
\int d^2x & \biggl[
 i\rho\partial_\mu\tilde{C}^\mu
 -i\lambda\epsilon^{\mu\nu}\partial_\mu\tilde{C}_\nu
 -i\chi_{\mu a}(D^\mu \tilde{C})^a
\biggr]\\
&=\int d^2x \biggl[
  i\text{Tr}(\overline{\psi}\gamma^\mu\partial_\mu\psi)
   +i\text{Tr}(\overline{\chi}\gamma^\mu \partial_\mu\chi)
  \biggr],
\end{split}
\end{align}
where $(\overline{\psi},\overline{\chi})=(C\psi^TC^{-1},C\chi^TC^{-1})
=(\psi^T,\chi^T)$.

Thus the action turns out to be the following action with the Dirac-K\"ahler
fermions $\psi$ and $\chi$:
\begin{align}
\begin{split}
S=\int d^2x \biggl[
 & \frac{1}{2}F^{\mu\nu}F_{\mu\nu}+(D_\mu\phi)_i(D^\mu\phi)^i+|\phi|^4 \\
 & +i\text{Tr}(\overline{\psi}\gamma^\mu\partial_\mu\psi)
   +i\text{Tr}(\overline{\chi}\gamma^\mu D_\mu\chi)\\
 & -4i\phi_1\text{Tr}(\overline{\psi}\gamma^5\chi)
  +4i\phi_1\text{Tr}(\overline{\psi}\gamma^5\chi\gamma^5)\\
 & -i\sqrt{2}A\text{Tr}(\overline{\chi}\gamma^5\chi)
  +i\sqrt{2}B\text{Tr}(\overline{\chi}\chi\gamma^5)\\
 & -\tfrac{1}{2}A\partial^2A+\tfrac{1}{2}B\partial^2B+2(A^2-B^2)|\phi|^2
\biggr],
\end{split}
\end{align}
where $D_\mu\chi=\partial_\mu\chi+2\omega_\mu \chi\gamma^5$, and
\begin{align}
\overline{\eta}= \tfrac{1}{2\sqrt{2}}(A-B),\
 \eta = \sqrt{2}(A+B).
\end{align}
This action is equivalent to N=2 super Yang-Mills with the Abelian Higgs 
system \cite{DVF} and also topological Bogomol'nyi theory 
\cite{SchapT} except 
for the symmetry breaking potential.

\section{Twisted N=D=2 superspace formalism}

\setcounter{equation}{0}

In this section we propose the twisted superspace formalism which will 
reproduce the off-shell N=D=2 BF model and the super Yang-Mills action of 
the previous section. 
\subsection{Twisted superspace and superfield}
We consider the following group element:
\begin{align}
G(x^\mu,\theta,\theta^\mu,\tilde{\theta})
 =e^{i(-x^\mu P_\mu + \theta Q +\theta^\mu Q_\mu +\tilde{\theta}\tilde{Q})},
\end{align}
where all $\theta$'s are anticommuting parameters. Twisted D=N=2 superspace
is defined in the parameter space of 
$(x^\mu,\theta,\theta^\mu,\tilde{\theta})$.

By using the relations (\ref{eq:N=2 T-algebra}) and the Hausdorff's formula
$e^Ae^B=e^{A+B+\frac{1}{2}[A,B]+\cdots}$, we can show the following
relation:
\begin{align}
G(0,\xi,\xi^\mu,\tilde{\xi})
G(x^\mu,\theta,\theta^\mu,\tilde{\theta})
=G(x^\mu+a^\mu,\theta+\xi,\theta^\mu+\xi^\mu,\tilde{\theta}+\tilde{\xi}),
\end{align}
where $a^\mu=\frac{i}{2}\xi\theta^\mu+\tfrac{i}{2}\xi^\mu\theta
 +\tfrac{i}{2}\epsilon^\mu{_\nu}\xi^\nu\tilde{\theta}
 +\tfrac{i}{2}\epsilon^\mu{_\nu}\tilde{\xi}\theta^\nu$.
This multiplication induces a shift transformation in superspace
$(x^\mu,\theta,\theta^\mu,\tilde{\theta})$:
\begin{align}
(x^\mu,\theta,\theta^\mu,\tilde{\theta})
\rightarrow
(x^\mu+a^\mu,\theta+\xi,\theta^\mu+\xi^\mu,\tilde{\theta}+\tilde{\xi}).
\end{align}
This transformation can be generated by the differential operators $Q$,
$Q_\mu$, and $\tilde{Q}$:
\begin{equation}
\begin{split}
Q
 &=\frac{\partial}{\partial \theta}
  +\frac{i}{2}\theta^\mu\partial_\mu,\\
Q_\mu
 &=\frac{\partial}{\partial \theta^\mu}
  +\frac{i}{2}\theta\partial_\mu
  -\frac{i}{2}\tilde{\theta}\epsilon_{\mu\nu}\partial^\nu,\\
\tilde{Q}
 &=\frac{\partial}{\partial \tilde{\theta}}
  -\frac{i}{2}\theta^\mu\epsilon_{\mu\nu}\partial^\nu.\\
\end{split}
\label{eq:Q-operator}
\end{equation}
Indeed we find
\begin{align}
\delta_\xi
\begin{pmatrix}
x^\mu\\
\theta\\
\theta^\mu\\
\tilde{\theta}
\end{pmatrix}
=(\xi Q+\xi^\mu Q_\mu+\tilde{\xi}\tilde{Q})
\begin{pmatrix}
x^\mu\\
\theta\\
\theta^\mu\\
\tilde{\theta}
\end{pmatrix}
=
\begin{pmatrix}
a^\mu\\
\xi\\
\xi^\mu\\
\tilde{\xi}
\end{pmatrix}
.
\end{align}
These operators satisfy the following relations:
\begin{equation}
\begin{split}
& \{Q,Q_\mu\}=i\partial_\mu,\
 \{\tilde{Q},Q_\mu\}=-i\epsilon_{\mu\nu}\partial^\nu,\\
& Q^2=\tilde{Q}^2=\{Q,\tilde{Q}\}=\{Q_\mu,Q_\nu\}=0.
\end{split}
\label{N=2 left operator algebra}
\end{equation}

The general scalar superfields in twisted D=N=2 superspace are defined as
the functions of $(x^\mu,\theta,\theta^\mu,\tilde{\theta})$, and can be
expanded as follows:
\begin{equation}
\begin{split}
F(x^\mu,\theta,\theta^\mu,\tilde{\theta})
 =&\phi(x)+\theta^\mu\phi_\mu(x)+\theta^2 \tilde{\phi}(x)\\
 &+\theta\Bigl( \psi(x)+\theta^\mu\psi_\mu(x)
   +\theta^2 \tilde{\psi}(x)\Bigr)\\
 &+\tilde{\theta}\Big(\chi(x)+\theta^\mu\chi_\mu(x)
   +\theta^2 \tilde{\chi}(x)\Bigr)\\
 &+\theta\tilde{\theta}\Bigl(\lambda(x)+\theta^\mu\lambda_\mu(x)
   +\theta^2 \tilde{\lambda}(x)\Bigr),\\
\end{split}
\label{eq:F}
\end{equation}
where the leading component $\phi(x)$ can be taken to be not only bosonic
but also fermionic.

The transformation law of the superfield $F$ is defined as follows:
\begin{equation}
\begin{split}
\delta_\xi F(x^\mu,\theta,\theta^\mu,\tilde{\theta})
=&\delta_\xi\phi(x)+\theta^\mu \delta_\xi\phi_\mu(x)+\theta^2
\delta_\xi\tilde{\phi}(x)\\
&+\theta\Bigl(\delta_\xi\psi(x)+\theta^\mu \delta_\xi\psi_\mu(x)
   +\theta^2 \delta_\xi\tilde{\psi}(x)\Bigr)\\
&+\tilde{\theta}\Big(\delta_\xi\chi(x)+\theta^\mu \delta_\xi\chi_\mu(x)
  +\theta^2 \delta_\xi\tilde{\chi}(x)\Bigr)\\
&+\theta\tilde{\theta}\Bigl(\delta_\xi\lambda(x)+\theta^\mu
\delta_\xi\lambda_\mu(x)
  +\theta^2 \delta_\xi\tilde{\lambda}(x)\Bigr)\\
=&(\xi Q+\xi^\mu
Q_\mu+\tilde{\xi}\tilde{Q})F(x^\mu,\theta,\theta^\mu,\tilde{\theta}),
\end{split}
\label{eq:F-trans}
\end{equation}
where $Q$, $Q_\mu$ and $\tilde{Q}$ are the differential operators
(\ref{eq:Q-operator}). The transformation laws of the component fields
$\phi^A(x)=(\phi(x),\phi_\mu(x),\tilde{\phi}(x),\dots)$ are obtained 
by comparing coefficients of the same superspace parameters in 
(\ref{eq:F-trans}) (see Table 5). Those
transformation laws lead to the following super-charge algebra:
\begin{equation}
\begin{split}
& \{Q,Q_\mu\}=-i\partial_\mu,\
 \{\tilde{Q},Q_\mu\}=i\epsilon_{\mu\nu}\partial^\nu,\\
& Q^2=\tilde{Q}^2=\{Q,\tilde{Q}\}=\{Q_\mu,Q_\nu\}=0,
\end{split}
\end{equation}
which are the same of (\ref{eq:N=2 T-algebra}) with $P_\mu=-i\partial_\mu$.
It should be noted that the only difference between the super charge 
algebra and the corresponding differential operator algebra is 
a sign difference for the derivative. 

\begin{table}[htbp]
\begin{center}
\begin{tabular}{c|c|c|c}
\hline
$\phi^A$ & $Q\phi^A$ & $Q_\mu\phi^A$ & $\tilde{Q}\phi^A$ \\
\hline
$\phi$
 & $\psi$
 & $\phi_\mu$
 & $\chi$ \\
$\phi_\rho$
 & $-\psi_\rho-\frac{i}{2}\partial_\rho\phi$
 & $-\epsilon_{\mu\rho}\tilde{\phi}$
 & $-\chi_\rho+\frac{i}{2}\epsilon_{\rho\sigma}\partial^\sigma\phi$ \\
$\tilde{\phi}$
 & $\tilde{\psi}+\frac{i}{2}\epsilon^{\rho\sigma}\partial_\rho\phi_\sigma$
 & $0$
 & $\tilde{\chi}-\frac{i}{2}\partial^\rho\phi_\rho$ \\
\hline
$\psi$
 & $0$
 & $\psi_\mu-\frac{i}{2}\partial_\mu\phi$
 & $\lambda$ \\
$\psi_\rho$
 & $-\frac{i}{2}\partial_\rho\psi$
 & $-\epsilon_{\mu\rho}\tilde{\psi}+\frac{i}{2}\partial_\mu\phi_\rho$
 & $-\lambda_\rho+\frac{i}{2}\epsilon_{\rho\sigma}\partial^\sigma\psi$ \\
$\tilde{\psi}$
 & $\frac{i}{2}\epsilon^{\rho\sigma}\partial_\rho\psi_\sigma$
 & $-\frac{i}{2}\partial_\mu\tilde{\phi}$
 & $\tilde{\lambda}-\frac{i}{2}\partial^\rho\psi_\rho$ \\
\hline
$\chi$
 & $-\lambda$
 & $\chi_\mu+\frac{i}{2}\epsilon_{\mu\nu}\partial^\nu\phi$
 & $0$ \\
$\chi_\rho$
 & $\lambda_\rho-\frac{i}{2}\partial_\rho\chi$
 & $-\epsilon_{\mu\rho}\tilde{\chi}-\frac{i}{2}
\epsilon_{\mu\nu}\partial^\nu \phi_\rho$
 & $+\frac{i}{2}\epsilon_{\rho\sigma}\partial^\sigma\chi$ \\
$\tilde{\chi}$
 & $-\tilde{\lambda}+\frac{i}{2}\epsilon^{\rho\sigma}\partial_\rho\chi_\sigma$
 & $\frac{i}{2}\epsilon_{\mu\nu}\partial^\nu\tilde{\phi}$
 & $-\frac{i}{2}\partial^\rho\chi_\rho$ \\
\hline
$\lambda$
 & $0$
 & $\lambda_\mu+\frac{i}{2}\partial_\mu\chi+
 \frac{i}{2}\epsilon_{\mu\nu}\partial^\nu\psi$
 & $0$ \\
$\lambda_\rho$
 & $-\frac{i}{2}\partial_\rho\lambda$
 & $-\epsilon_{\mu\rho}\tilde{\lambda}
  -\frac{i}{2}\partial_\mu\chi_\rho
  -\frac{i}{2}\epsilon_{\mu\nu}\partial^\nu\psi_\rho$
 & $\frac{i}{2}\epsilon_{\rho\sigma}\partial^\sigma\lambda$ \\
$\tilde{\lambda}$
 & $\frac{i}{2}\epsilon^{\rho\sigma}\partial_\rho\lambda_\sigma$
 & $\frac{i}{2}\partial_\mu\tilde{\chi}
  +\frac{i}{2}\epsilon_{\mu\nu}\partial^\nu\tilde{\psi}$
 & $-\frac{i}{2}\partial^\rho\lambda_\rho$ \\
\hline
\end{tabular}
\caption{Transformation laws of the component fields of $F$}
\end{center}
\end{table}

Given the transformation laws of the component fields, we can expand the
superfield $F$ as follows:
\begin{equation}
\begin{split}
F(x^\mu,\theta,\theta^\mu,\tilde{\theta})
 &=e^{\delta_\theta}\phi(x)\\
 &=\phi(x)+\delta_\theta\phi(x)
  +\frac{1}{2}\delta_\theta{^2}\phi(x)
  +\frac{1}{3!}\delta_\theta{^3}\phi(x)
  +\frac{1}{4!}\delta_\theta{^4}\phi(x),
\end{split}
\end{equation}
where $\delta_\theta$ is defined as
\begin{equation}
\delta_\theta=\theta Q+\theta^\mu Q_\mu+\tilde{\theta}\tilde{Q}.
\label{def_delta_theta}
\end{equation}

As we have seen, the differential operators in (\ref{eq:Q-operator}) generate
the shift transformation of superspace induced by left multiplication
$G(0,\xi,\xi^\mu,\tilde{\xi})G(x^\mu,\theta,\theta^\mu,\tilde{\theta})$. On
the other hand, there exist the differential operators which generate the
shift transformation induced by right multiplication
$G(x^\mu,\theta,\theta^\mu,\tilde{\theta})G(0,\xi,\xi^\mu,\tilde{\xi})$:
\begin{equation}
\begin{split}
D
 &=\frac{\partial}{\partial \theta}
  -\frac{i}{2}\theta^\mu\partial_\mu,\\
D_\mu
 &=\frac{\partial}{\partial \theta^\mu}
  -\frac{i}{2}\theta\partial_\mu
  +\frac{i}{2}\tilde{\theta}\epsilon_{\mu\nu}\partial^\nu,\\
\tilde{D}
 &=\frac{\partial}{\partial \tilde{\theta}}
  +\frac{i}{2}\theta^\mu\epsilon_{\mu\nu}\partial^\nu ,
\end{split}
\label{eq:D-operator}
\end{equation}
which satisfy the relations:
\begin{equation}
\begin{split}
\{D,D_\mu\}&=-i\partial_\mu,\
\{\tilde{D},D_\mu\}=i\epsilon_{\mu\nu}\partial^\nu,\\
D^2&=\tilde{D}^2=\{D,\tilde{D}\}=\{D_\mu,D_\nu\}=0,\\
\end{split}
\label{N=2 right operator algebra}
\end{equation}
where only the sign of $\partial_\mu$ is changed from 
the left operator algebra (\ref{N=2 left operator algebra}).
$Q^A=(Q,Q_\mu,\tilde{Q})$ and $D^A=(D,D_\mu,\tilde{D})$ anticommute:
\begin{align}
\{Q^A,D^B\}=0.
\end{align}

\subsection{Chiral decomposition of twisted supercharge}
We have defined the general scalar superfields in the previous subsection. 
However their representations are reducible. We can obtain the irreducible
representations by imposing chiral conditions. In this subsection we introduce
the twisted chiral differential operators by which superfields can be
classified.

Let us first introduce the following chiral generators:
\begin{align}
Q^\pm
 = \frac{1}{\sqrt{2}}(Q\mp i\tilde{Q}),
\quad
{Q^\pm}_\mu
 = \frac{1}{\sqrt{2}}(Q_\mu\pm i\epsilon_{\mu\nu}Q^\nu).
\label{twisted_chiral_generators} 
\end{align}
As we can see from (\ref{eq:N=2 T-algebra}), these new generators satisfy the
following relations:
\begin{align}
Q^{\pm 2}&=0,
 & \{Q^+,Q^-\}&=0, \nonumber\\
\{Q^\pm,{Q^\pm}_\mu\}&=\delta_{\mu\nu}^\pm P^\nu,
 & \{Q^\pm,{Q^\mp}_\mu\}&=0,\\
\{{Q^\pm}_\mu,{Q^\pm}_\nu\}&=0,
 & \{{Q^+}_\mu,{Q^-}_\nu\}&=0,
 \nonumber
\end{align}
where
\begin{equation}
\delta_{\mu\nu}^\pm = \delta_{\mu\nu} \pm i\epsilon_{\mu\nu}.
\label{delta_pm}
\end{equation}
The super-transformation $\delta_\theta$ can be represented by the
generators $Q^+$, ${Q^+}_\mu$, $Q^-$, and ${Q^-}_\mu$:
\begin{align}
\delta_\theta=\theta Q+\theta^\mu Q_\mu+\tilde{\theta}\tilde{Q}
=\theta^- Q^+ +\tfrac{1}{2}\theta^{-\mu} {Q^+}_\mu
+\theta^+ Q^- +\tfrac{1}{2}\theta^{+\mu} {Q^-}_\mu,
\label{delta_theta}
\end{align}
where
\begin{equation}
\theta^\pm
 = \frac{1}{\sqrt{2}}(\theta\mp i\tilde{\theta}),
\quad
{\theta^\pm}_\mu
 = \frac{1}{\sqrt{2}}(\theta_\mu\pm i\epsilon_{\mu\nu}\theta^\nu).
\label{eq:theta+-}
\end{equation}
These new parameters satisfy the following properties:
\begin{equation}
(\theta^\pm)^*=\theta^\mp,
\ \ \ \ 
({\theta^\pm}_\mu)^*={\theta^\mp}_\mu, 
\end{equation}
where $\theta^{+\mu}$ and $\theta^{-\mu}$ are self-dual and anti-self-dual 
parameters, respectively, in the following sense:
\begin{align}
\theta^{\pm\mu}=\pm i\epsilon^{\mu\nu}{\theta^\pm}_\nu.
\end{align}

Now we redefine superspace by newly defined fermionic chiral parameters \\
$(x^\mu,\theta^+,\theta^{+\mu},\theta^-,\theta^{-\mu})$. 
The differential operators corresponding to the chiral generators 
defined in (\ref{twisted_chiral_generators}) are obtained from 
(\ref{eq:Q-operator})
\begin{equation}
\begin{split}
Q^\mp
 &= \frac{\partial}{\partial \theta^\pm}
  +\frac{i}{4}\theta^{\pm\mu}{\partial^\mp}_\mu , \\
{Q^\mp}_\mu
 &= \frac{\partial}{\partial \theta^{\pm\mu}}
  +\frac{i}{2}\theta^{\pm}{\partial^\mp}_\mu ,
\end{split}
\label{eq:Q+-}
\end{equation}
which satisfy the following operator relations:
\begin{align}
Q^{\pm 2}&=0,
 & \{Q^+,Q^-\}&=0, \nonumber\\
\{Q^\pm,{Q^\pm}_\mu\}&=i{\partial^\pm}_\mu,
 & \{Q^\pm,{Q^\mp}_\mu\}&=0,\\
\{{Q^\pm}_\mu,{Q^\pm}_\nu\}&=0,
 & \{{Q^+}_\mu,{Q^-}_\nu\}&=0,
 \nonumber
\end{align}
with  
\begin{equation}
{\partial^\pm}_\mu=\partial_\mu\pm i\epsilon_{\mu\nu}\partial^\nu.
\label{partial_pm} 
\end{equation}
It is worth to note the following relations with the current conventions:
\begin{eqnarray}
(Q^\mp)^* &=& -Q^\pm, \ \ \ (Q^\mp_\mu)^* = -Q^\pm_\mu \nonumber \\
\left(\frac{\partial}{\partial \theta^\pm}\right)^* &=& 
- \frac{\partial}{\partial \theta^\mp}, \ \ \  
\left(\frac{\partial}{\partial \theta^{\pm\mu}}\right)^* = 
-\frac{\partial}{\partial \theta^{\mp\mu}} \nonumber \\
\partial^\pm_\mu x^\nu &=& 
\frac{\partial}{\partial \theta^{\pm \mu}} \theta^{\pm\nu} = 
{\delta^{\pm}{}_{\mu}}^\nu,
\end{eqnarray}
where $\delta_{\mu \nu}^{\pm}$ is defined in (\ref{delta_pm}).

The differential operators which generate the shift from right multiplication 
are obtained from (\ref{eq:D-operator}) 
\begin{equation}
\begin{split}
D^\mp
 &\equiv \frac{1}{\sqrt{2}}(D\pm i\tilde{D})
 = \frac{\partial}{\partial \theta^\pm}
  -\frac{i}{4}\theta^{\pm\mu}{\partial^\mp}_\mu , \\
{D^\mp}_\mu
 &\equiv \frac{1}{\sqrt{2}}(D_\mu\mp i\epsilon_{\mu\nu}D^\nu)
 = \frac{\partial}{\partial \theta^{\pm\mu}}
  -\frac{i}{2}\theta^{\pm}{\partial^\mp}_\mu ,
\end{split}
\label{eq:D+-}
\end{equation}
which satisfy
\begin{align}
D^{\pm 2}&=0,
 & \{D^+,D^-\}&=0, \nonumber\\
\{D^\pm,{D^\pm}_\mu\}&=-i{\partial^\pm}_\mu,
 & \{D^\pm,{D^\mp}_\mu\}&=0,\\
\{{D^\pm}_\mu,{D^\pm}_\nu\}&=0,
 & \{{D^+}_\mu,{D^-}_\nu\}&=0.
 \nonumber
\end{align}
The differential operators $Q^{\prime A}=(Q^\pm,Q^{\pm\mu})$ and 
$D^{\prime A}=(D^\pm,D^{\pm\mu})$ anticommute:
\begin{align}
\{Q^{\prime A},D^{\prime B}\}=0.
\end{align}

\subsection{Chiral superfields}
We can classify the chiral superfields into four types which are
constrained by the following four types of conditions:
\begin{align}
D^+\Psi&={D^-}\Psi=0,
 \label{eq:chiral 1}\\
{D^+}_\mu\overline{\Psi}&={D^-}_\mu\overline{\Psi}=0,
 \label{eq:chiral 2}\\
{D^+}_\mu\Phi&=D^-\Phi=0,
 \label{eq:chiral 3}\\
D^+\Phi^+&={D^-}_\mu\Phi^+=0.
 \label{eq:chiral 4}
\end{align}
Here in this subsection we consider that the superfield has single component 
and thus all the component fields have abelian nature. 

\subsubsection{$\Psi$ and $\overline{\Psi}$} 
Firstly, we consider the chiral superfields characterized by the conditions
(\ref{eq:chiral 1}) and (\ref{eq:chiral 2}).

From the definition (\ref{eq:D+-}) we see that the condition
(\ref{eq:chiral 1}) is equivalent to the condition
\begin{align}
D\Psi=\tilde{D}\Psi=0.
\label{scalar condition}
\end{align}
The superfields satisfying the above condition are functions of
\begin{equation}
z^\mu=x^\mu+\frac{i}{2}\theta\theta^\mu-\frac{i}{2}\epsilon^\mu{_\nu}\theta
^\nu\tilde{\theta} 
\end{equation}
and $\theta^\mu$ since 
\begin{equation}
Dz^\mu=D\theta^\mu=0, \ \ \ \tilde{D}z^\mu=\tilde{D}\theta^\mu=0.
\end{equation}
In general we have the following relation:
\begin{eqnarray}
e^{\theta^\mu Q_\mu} e^{\theta Q + \tilde{\theta}\tilde{Q}}\varphi(x^\mu) 
&=& e^{\delta_\theta - \frac{1}{2}\theta\theta^\mu i\partial_\mu + 
\frac{1}{2}\tilde{\theta}\theta^\mu i\epsilon_{\mu\nu}\partial^\nu}
\varphi(x^\mu) 
\nonumber \\
&=&e^{\delta_\theta}\varphi\left(x^\mu - \frac{i}{2}\theta\theta^\mu  + 
\frac{i}{2}\epsilon^{\mu\nu}\theta_\nu\tilde{\theta}\right),
\end{eqnarray}
and thus
\begin{equation}
e^{\delta_\theta}\varphi(x^\mu) = e^{\theta^\mu Q_\mu} 
e^{\theta Q+ \tilde{\theta}\tilde{Q}}\varphi(z^\mu),
\label{general_relation}
\end{equation}
where $\delta_\theta$ is given by (\ref{def_delta_theta}).
Therefore the chiral superfield $\Psi$ can be expanded as follows:
\begin{equation}
\begin{split}
\Psi&=\Psi(z^\mu,\theta^\mu)
 = e^{\delta_\theta}\phi(x)
 =e^{\theta^\mu Q_\mu}\phi(z) \\
 &=\phi(z)+\theta^\mu \phi_\mu(z) +\theta^2 \tilde{\phi}(z)\\
 &=\phi(x)+\theta^\mu\phi_\mu(x)\\
 &\quad
  +\tfrac{i}{2}\theta\theta^\mu \partial_\mu \phi(x)
  +\theta^2\tilde{\phi}(x)
  -\tfrac{i}{2}\epsilon^\mu{_\nu}\theta^\nu\tilde{\theta}\partial_\mu
\phi(x)\\
 &\quad
  +\tfrac{i}{2}\theta\theta^2\epsilon^{\mu\nu}\partial_\mu\phi_\nu
  -\tfrac{i}{2}\theta^2\tilde{\theta}\partial^\mu\phi_\mu
  +\tfrac{1}{4}\theta^4\partial^2 \phi(x),
\label{psi_superfield}  
\end{split}
\end{equation}
where $\theta^2=\frac{1}{2}\epsilon_{\mu\nu}\theta^\mu\theta^\nu$ 
and $\theta^4=\theta\tilde{\theta}\theta^2$.
The set $(\phi,\phi_\mu,\tilde{\phi})$ constructs twisted N=2 off-shell 
super multiplet(see Table 6).

\begin{table}[htbp]
\begin{center}
\begin{tabular}{c|c|c|c}
\hline
$\phi^A$ & $Q\phi^A$ & $Q_\mu\phi^A$ & $\tilde{Q}\phi^A$\\
\hline
$\phi$ & $0$ & $\phi_\mu$ & $0$\\
$\phi_\rho$ & $-i\partial_\rho \phi$ & $-\epsilon_{\mu\rho}\tilde{\phi}$
  & $i\epsilon_{\rho\nu}\partial^\nu \phi$\\
$\tilde{\phi}$ & $i\epsilon^{\rho\sigma}\partial_\rho\phi_\sigma$
  & $0$ & $-i\partial^\rho \phi_\rho$\\
\hline
$\psi$ & $\chi$ & $0$ & $\tilde{\chi}$\\
$\chi$ & $0$ & $-i\partial_\mu\psi$ & $\tilde{\psi}$\\
$\tilde{\chi}$ & $-\tilde{\psi}$ & $i\epsilon_{\mu\nu}\partial^\nu \psi$ &
$0$\\
$\tilde{\psi}$ & $0$
 & $i\epsilon_{\mu\nu}\partial^\nu \chi+i\partial_\mu\tilde{\chi}$ & $0$\\  
\hline
\end{tabular}
\caption{Chiral twisted super multiplet $(\phi,\phi_\mu,\tilde{\phi})$ and 
$(\psi,\chi,\tilde{\chi},\tilde{\psi})$.}
\end{center}
\end{table}

On the other hand we can see from (\ref{eq:D+-}), 
the conditions in (\ref{eq:chiral 2}) are equivalent to the
condition
\begin{equation}
D_\mu\overline{\Psi}=0.
\label{anti-chiral-condition}
\end{equation}
This type of chiral superfields are the functions of 
\begin{equation}
\tilde{z}^\mu=x^\mu-\frac{i}{2}\theta\theta^\mu+\frac{i}{2}
\epsilon^\mu{}_\nu\theta^\nu\tilde{\theta},
\end{equation}
and $\theta$, $\tilde{\theta}$ 
since 
\begin{equation}
D_\mu\tilde{z}^\nu=0, \ \ \ D_\mu\theta=D_\mu\tilde{\theta}=0.
\end{equation}
Then $\overline{\Psi}$ can be expanded as follows:
\begin{equation}
\begin{split}
\overline{\Psi}
 &=\overline{\Psi}(\tilde{z}^\mu,\theta,\tilde{\theta})
 = e^{\delta_\theta}\psi(x)
 =e^{\theta Q+\tilde{\theta}\tilde{Q}}\psi(\tilde{z}) \\
 &=\psi(\tilde{z})+\theta \chi(\tilde{z})
  +\tilde{\theta}\tilde{\chi}(\tilde{z})
  +\theta \tilde{\theta}\tilde{\psi}(\tilde{z})\\
 &=\psi(x)+\theta\chi+\tilde{\theta}\tilde{\chi}\\
 &\quad
  +\theta\tilde{\theta}\tilde{\psi}(x)
  -\tfrac{i}{2}\theta\theta^\mu \partial_\mu \psi(x)
  +\tfrac{i}{2}\epsilon^\mu{_\nu}\theta^\nu\tilde{\theta}\partial_\mu
\psi(x)\\
 &\quad
  -\tfrac{i}{2}\theta\theta^\mu\tilde{\theta}
   (\epsilon_{\mu\nu}\partial^\nu\chi+\partial_\mu\tilde{\chi})
  +\tfrac{1}{4}\theta^4\partial^2 \psi(x).
\label{psibar_superfield}
\end{split}
\end{equation}
The set $(\psi,\chi,\tilde{\chi},\tilde{\psi})$ also constructs a super
multiplet(see Table 6).


We now introduce off-shell N=2 supersymmetric action:
\begin{align}
S=\int d^2x \int d^4\theta
 \Bigl(i^{\epsilon_\Psi} \overline\Psi(x^\mu,\theta,\theta^\mu,\tilde{\theta})
 {\Psi}(x^\mu,\theta,\theta^\mu,\tilde{\theta})
 \Bigr).
\label{eq:WZ action}
\end{align}
We can take the chiral superfields to be not only bosonic but also
fermionic. $\epsilon_\Psi$ should be taken $0$ or $1$ for bosonic or
fermionic $(\Psi,\overline{\Psi})$, respectively.

For fermionic $(\Psi,\overline{\Psi})$, the fields in the expansion 
of the superfield (\ref{psi_superfield}) and (\ref{psibar_superfield}) 
can be renamed as:
\begin{equation}
\begin{split}
\Psi&=\Psi(z^\mu,\theta^\mu)
 =ie^{\theta^\mu Q_\mu}c(z) 
 =ic(z)+\theta^\mu \omega_\mu(z) +i\theta^2 \lambda(z) \\
\overline{\Psi}
 &=\overline{\Psi}(\tilde{z}^\mu,\theta,\tilde{\theta})
  =ie^{\theta Q + \tilde{\theta} \tilde{Q}}\overline{c}(\tilde{z})
  =i\overline{c}(\tilde{z})+\theta b(\tilde{z})
  +\tilde{\theta}\phi(\tilde{z})
  -i\theta \tilde{\theta}\rho(\tilde{z}),
\label{expansion_psi_psibar}  
\end{split}
\end{equation}
where we have the correspondence of the fields; 
$(\psi,\chi,\tilde{\chi},\tilde{\psi})\rightarrow 
(i\overline{c},b,\phi,-i\rho)$ and
$(\phi,\phi_\mu,\tilde{\phi})\rightarrow (ic,\omega_\mu,i\lambda)$.
Then the action (\ref{eq:WZ action}) leads
\begin{align}
S_f&= \int d^2x \int d^4\theta \  
 ( i \overline{\Psi} \Psi )
 = \int d^2x \int d^4\theta \ e^{\delta_\theta}(-i\overline{c}c)
 \nonumber \\
 &=\int_{M_2} d^2x \ 
Q\tilde{Q}\frac{1}{2}\epsilon^{\mu\nu}Q_\mu Q_\nu (-i\bar{c}c)\nonumber\\
 &=\int d^2x
\Bigl(
\phi\epsilon^{\mu\nu}\partial_\mu\omega_\nu
+b \partial^\mu\omega_\mu
+i\partial_\mu\overline{c}\partial^\mu c
-i\lambda\rho
\Bigr),
\label{eq:WZ 3}
\end{align}
where $\delta_\theta$ is defined in (\ref{def_delta_theta}).
This action is exactly the same as the quantized BF action 
(\ref{eq:QBF action2}) which is off-shell N=2 supersymmetric. 
The off-shell N=2 supersymmetry transformations of the fields 
in (\ref{expansion_psi_psibar}) is given in Table 6 and is exactly 
the same as those of Table 2. 

As we have already pointed out that the simple form of the action 
(\ref{eq:QBF action3}) has suggested a hidden mechanism of a superspace 
formalism. We have in fact found the twisted N=2 superspace formalism 
which reproduces the quantized BF action up to the last term 
$-i\lambda\rho$ which includes auxiliary fields and is crucial to 
fufill off-shell N=2 supersymmetry. 
We find the following correspondence:
\begin{equation}
s=Q, \ \ \ s_\mu = Q_\mu, \ \ \ \tilde{s} = \tilde{Q}. 
\end{equation}
where $s=Q$ can be regarded as BRST charge associated with the gauge 
symmetry $\delta_{gauge}\omega_\mu=\partial_\mu v$ with the Landau type
gauge fixing, $\partial_\mu\omega^\mu=0$. Integrating $\lambda$ and 
$\rho$ out, 
the action (\ref{eq:WZ 3}) coincides exactly with the quantized BF 
action (\ref{eq:QBF action}), where the off-shell N=2 supersymmetry reduces 
to on-shell N=2 supersymmetry.

For bosonic $(\Psi,\overline{\Psi})$ the action (\ref{eq:WZ action}) can be
written as follows:
\begin{align}
S_b&= \int d^2x \int d^4\theta \  
 (  \overline{\Psi} \Psi )
 = \int d^2x \int d^4\theta \ e^{\delta_\theta}(\psi\phi)
 \nonumber \\
 &=\int_{M_2} d^2x \ 
Q\tilde{Q}\frac{1}{2}\epsilon^{\mu\nu}Q_\mu Q_\nu (\psi\phi)\nonumber\\
 &=\int d^2x
\Bigl(
i\tilde{\chi}\epsilon^{\mu\nu}\partial_\mu\phi_\nu
+i\chi\partial^\mu\phi_\mu
-\partial^\mu\psi\partial_\mu\phi
+\tilde{\psi}\tilde{\phi}
\Bigr),
\label{eq:WZ action 2}
\end{align}
where $(\phi,\psi,\tilde{\phi},\tilde{\psi})$ and
$(\phi_\mu,\chi,\tilde{\chi})$ are bosonic and fermionic fields,
respectively.
The fermionic terms in (\ref{eq:WZ action 2}) change into matter fermions via 
Dirac-K\"ahler fermion mechanism:
\begin{align}
\int d^2x\ \left(
  i\tilde{\chi} \epsilon^{\mu\nu}\partial_\mu\phi_\nu 
  +i\chi \partial^\mu\phi_\mu 
 \right)
=\int d^2x\
  \text{Tr} \left(
   i\overline{\xi} \gamma^\mu \partial_\mu \xi
  \right)
\end{align}
where the Dirac-K\"ahler fermion $\xi$ is defined as
\begin{align}
\xi_{\alpha\beta}
=\frac{1}{2}\left(
 \mathbf{1}\chi+\gamma^\mu\phi_\mu+\gamma^5\tilde{\chi}
\right)_{\alpha\beta},
\label{DKtr1}
\end{align}
with $\overline{\xi}=C\xi^TC=\xi^T$. 
We can recognize that each spinor suffix of this Dirac-K\"ahler fermion 
has the Majorana Weyl fermion nature. 
We now redefine the bosonic fields as
follows:
\begin{equation}
\begin{split}
\phi_0 &=\frac{1}{2}(\phi+\psi),
 \phi_1 =\frac{1}{2}(\phi-\psi),\\
F_0&=\frac{1}{2}(\tilde{\phi}+\tilde{\psi}),
 F_1 =\frac{1}{2}(\tilde{\phi}-\tilde{\psi}).
\end{split}
\end{equation}
Then the action (\ref{eq:WZ action 2}) can be rewritten by the new fields:
\begin{align}
S_b&=\int d^2x \sum_{i=1}^2
\Bigl(
 i{\xi_\alpha}^i {\gamma^\mu}_{\alpha\beta} \partial_\mu {\xi_\beta}^i
 +\partial_\mu\phi^i\partial^\mu\phi^i
 -F^iF^i
\Bigr),
\end{align}
where we further redefine 
$i\phi_0\rightarrow\phi_2$ and $iF_0\rightarrow F_2$
\footnote{In general the ghosts carry the indefinite metric and thus have 
non-unitary nature while this type of supersymmetric model is unitary. 
This change of sings for the quadratic terms can be related with 
this fact\cite{NakaK}. }.
This is the 2-dimensional version of N=2 Wess-Zumino action which 
has off-shell N=2 supersymmetry invariance. 
It is important to recognize at this stage that the second suffix 
of the Dirac-K\"ahler matter fermion in the action is N=2 extended 
supersymmetry suffix. 

This shows that the fermionic and bosonic chiral bi-linear form of the 
superfield actions lead an extended version of the quantized BF action and 
Wess-Zumino action, respectively, which have off-shell N=2 supersymmetry 
invariance.  
It is interesting to note that this extended quantized BF action 
(\ref{eq:WZ 3}) can be transformed to the D=N=2 Wess-Zumino action. 
To see this, we make a chain of redefinitions of fields, 
$\{\phi,\tilde{\phi}\} \rightarrow\{\phi,\tilde{\rho}\}
\rightarrow\{\psi_\mu\}$ and $\{\chi,\phi_\mu,\tilde{\chi}\}\rightarrow
\{\chi,\alpha,\beta,\tilde{\chi}\}\rightarrow
\{\chi,f,\beta,\tilde{\chi}\}\rightarrow \{\phi_0,\phi_1,F_0,F_1\}$:
\begin{equation}
\begin{split}
\tilde{\phi}&=-\partial^2\tilde{\rho},\
 \psi_\mu =\partial_\mu\phi+\epsilon_{\mu\nu}\partial^\nu\tilde{\rho},\\
\phi_\mu &= \partial_\mu \alpha+\epsilon_{\mu\nu}\partial^\nu\beta,\
 f=\partial^2\alpha,\
\left\{
\begin{array}{l}
\tilde{\chi}=\phi_0+\phi_1\\
\beta=\phi_0-\phi_1
\end{array}
\right.,\
\left\{
\begin{array}{l}
\chi=F_0+F_1\\
f=F_0-F_1.
\end{array}
\right.
\end{split}
\end{equation}
Then the action (\ref{eq:WZ 3}) can be rewritten as the following
Wess-Zumino action:
\begin{align}
S_f&=\int d^2x \sum_{i=1}^2
\Bigl(
 i{\psi_\alpha}^i {\gamma^\mu}_{\alpha\beta} \partial_\mu {\psi_\beta}^i
 +\partial_\mu\phi^i\partial^\mu\phi^i
 -F^iF^i
\Bigr),
\end{align}
where we again redefine $i\phi_0\rightarrow \phi_2$ and 
$iF_0\rightarrow F_1$, 
and the Dirac-K\"ahler matter fermion $\psi_{\alpha\beta}$ is defined as
\begin{align}
\psi_{\alpha\beta}
=\frac{1}{2}\left(
 \mathbf{1}\psi+\gamma^\mu\psi_\mu+\gamma^5\tilde{\psi}
\right)_{\alpha\beta}, 
\label{DKtr2}
\end{align}
where we have identified the second suffix of the Dirac-K\"ahler fermion 
as N=2 supersymmetry suffix. 

\subsubsection {Non-abelian Extension}

In this subsection we extend the abelian version of twisted superspace 
formulation of the previous subsection into non-abelian case. We point 
out that the extension from the abelian to a non-abelian version is 
straightforward, however, the chiral structure is sacrificed by the 
non-abelian nature.  
Using the general relation (\ref{general_relation}), we obtain the 
following superfield generated from the non-abelian ghost field:
\begin{eqnarray} 
\Psi &=& e^{\delta_\theta} c(x^\mu) = e^{\theta^\mu s_\mu} 
e^{\theta s + \tilde{\theta} \tilde{s}} c(z^\mu) \\
&=& e^{\theta^\mu s_\mu} (c + \theta (-c^2) ),
\end{eqnarray}
where we have introduced BRST transformation of the non-abelian ghost field: 
$sc = -c^2$ in Table 3. Thus the superfield $\Psi$ is the function of 
$z^\mu, \theta_\mu$ and $\theta$, and thus $\Psi$ is not twisted chiral 
superfield anymore since it includes $\theta$ and thus
\begin{align}
D\Psi \ne 0, \ \ \ \tilde{D}\Psi=0.
\label{scalar condition2}
\end{align}
We can yet expand the superfield $\Psi$ in the following form:
\begin{eqnarray} 
\Psi &=& c(z) - i \theta^\mu \omega_\mu(z) + \theta^2 \lambda(z) 
         - \theta c^2(z) + i \theta \theta^\mu [\omega_\mu(z),c(z)] 
\nonumber \\         
     & &    - \theta \theta^2 \left( \{\lambda(z),c(z)\} - 
         \frac{1}{2}\epsilon^{\mu\nu}[\omega_\mu(z),\omega_\nu(z)] \right)
\nonumber \\         
     &=&  c(x) + \frac{i}{2}\theta\theta^\mu\partial_\mu c(x)
     + \frac{i}{2}\theta^\mu\tilde{\theta}\epsilon_{\mu\nu}\partial^\nu c(x)
+ \frac{1}{4} \theta^4 \partial^2 c(x) - \theta c^2(x) 
- \frac{i}{2} \theta\theta^\mu \tilde{\theta} \epsilon_{\mu\nu}
\partial^\nu c^2(x)  \nonumber \\
& &  - i \theta^\mu \omega_\mu(x)
- \frac{1}{2} \theta^2 \tilde{\theta} \partial^\mu \omega_\mu(x) 
+\frac{1}{2} \theta \theta^2 \epsilon^{\mu\nu} \partial_\mu\omega_\nu(x) 
+ \theta^2 \lambda(x) 
+ i \theta \theta^\mu [\omega_\mu(x),c(x)] 
\nonumber \\
& &
+ \frac{1}{2} \theta^4 \partial^\mu [\omega_\mu(x),c(x)] 
- \theta \theta^2(\{\lambda(x),c(x)\} - \epsilon^{\mu\nu}[\omega_\mu(x),\omega_\nu(x)]).
\end{eqnarray} 
Similarly we can construct the superfield $\overline{\Psi}$ generated from 
the nob-abelian anti-ghost field:
\begin{eqnarray} 
\overline{\Psi} = e^{\delta_\theta} \overline{c}(x^\mu) = 
e^{\theta s + \tilde{\theta} \tilde{s}} \ e^{\theta^\mu s_\mu} \ 
\overline{c}(\tilde{z}^\mu) = 
e^{\theta s + \tilde{\theta}\tilde{s}} \ \overline{c}(\tilde{z}^\mu), 
\end{eqnarray}
which satisfies twisted anti-chiral condition (\ref{anti-chiral-condition}). 
We can expand the superfield $\overline{\Psi}$ in the following form:
\begin{eqnarray} 
\overline{\Psi} &=& \overline{c}(z) - i \theta b(z) - i \tilde{\theta}\phi(z) 
 + i \theta\tilde{\theta} \ ([\phi(z),c(z)] + i\rho(z)) \nonumber \\
&=& \overline{c}(x)
- \frac{i}{2} \theta \theta^\mu \partial_\mu \overline{c}(x) 
- \frac{i}{2} \theta^\mu \tilde{\theta} \epsilon_{\mu\nu}
\partial^\nu \overline{c}(x)
+ \frac{1}{4} \theta^4 \partial^2 \overline{c}(x)
- i \theta b(x) 
- \frac{1}{2} \theta \theta^\mu \tilde{\theta} 
\epsilon_{\mu\nu}\partial^\nu b(x) 
    \nonumber \\
& &  - i \tilde{\theta}\phi(x) 
- \frac{1}{2} \theta \theta^\mu \tilde{\theta} \partial_\mu \phi(x) 
+ i \theta \tilde{\theta} \ ([\phi(x),c(x)] + i\rho(x)).
\end{eqnarray}
Even though $\Psi$ is not twisted chiral superfield we can still 
construct the off-shell N=2 twisted super symmetric non-abelian BF 
action just like the abelian case:
\begin{align}
S_{\hbox{off-shell NABQBF}}&= \int d^2x \int d^4\theta \  
 \hbox{Tr}( i \overline{\Psi} \Psi )
 = \int d^2x \int d^4\theta \ e^{\delta_\theta}\hbox{Tr}(-i\overline{c}c)
 \nonumber \\
 &=\int d^2x \ 
s\tilde{s}\frac{1}{2}\epsilon^{\mu\nu}s_\mu s_\nu 
\hbox{Tr}(-i\bar{c}c)\nonumber\\
 &=\int d^2x 
  \ \hbox{Tr}
 [\phi F
 +b\partial^\mu\omega_\mu
 +i\partial^\mu \bar{c} D_\mu c -i\lambda\rho].
\label{eq:WZ 4}
\end{align}

\subsubsection{$\Phi$ and $\Phi^+$}
We consider chiral superfields constrained by the conditions 
(\ref{eq:chiral 3})
\begin{align}
D^-\Phi={D^+}_\mu\Phi=0,
\end{align}
where the gauge algebra is not introduced here. 
This type of chiral superfields are the functions of
$y^\mu=x^\mu+\frac{i}{2}\theta^+\theta^{+\mu}
-\frac{i}{2}\theta^-\theta^{-\mu}$, $\theta^{+\mu}$, and $\theta^-$, 
since 
\begin{eqnarray}
D^-y^\mu &=& D^-\theta^{+\mu} = D^-\theta^- =0, \nonumber \\
D_\mu^+y^\nu &=& D_\mu^+\theta^{+\nu} = D_\mu^+\theta^- = 0. 
\end{eqnarray}
Then the superfield can be expanded as follows:
\begin{equation}
\begin{split}
\Phi&=\Phi(y^\mu,\theta^{+\mu},\theta^-)\\
&=e^{\theta^- Q^+ + \frac{1}{2}\theta^{+\mu} Q_{\mu}^-}\phi(y)\\
&=\phi(y)
 +\frac{1}{2}\theta^{+\mu}{\xi^-}_\mu(y)
 +\theta^- \left(
  \xi(y)
  +\frac{1}{2}\theta^{+\mu}{K^-}_\mu(y)
 \right)\\
&=\phi(x)
 +\frac{1}{2}\theta^{+\mu}{\xi^-}_\mu(x)
 +\theta^-\xi(x)\\
& \quad
 +\frac{i}{4}\theta^+\theta^{+\mu}{\partial^-}_\mu\phi(x)
 -\frac{i}{4}\theta^-\theta^{-\mu}{\partial^+}_\mu\phi(x)
 +\frac{1}{2}\theta^-\theta^{+\mu}{K^-}_\mu(x)\\
& \quad 
+\frac{i}{8}\theta^-\theta^{+\rho}
{\theta^-}_\rho{\partial^{+\mu}}\xi^-_\mu (x)
 +\frac{i}{4}\theta^+\theta^{+\mu}\theta^-{\partial^-}_\mu\xi(x)\\
& \quad
 -\frac{1}{8}\theta^+\theta^-\theta^{+\mu}{\theta^-}_\mu \partial^2\phi(x),
\end{split}
\label{eq:Phi}
\end{equation}
where ${\xi^-}_\mu$ and ${K^-}_\mu$ are anti-self-dual. The N=2 supersymmetry 
transformation for the component fields $\phi$, ${\xi^-}_\mu$, $\xi$, 
and ${K^-}_\mu$ can be obtained by operating the differential operators 
(\ref{eq:Q+-}) to this representation (see Table 7).

Another type of chiral super fields $\Phi^+$ can be obtained by taking a 
conjugation of $\Phi$,
\begin{equation}
\begin{split}
\Phi^+ &= \Phi^* = \phi^*(x)
 -\frac{1}{2}\theta^{-\mu}{\xi^+}_\mu(x)
 -\theta^+\xi^*(x)\\
& \quad -\frac{i}{4}\theta^+\theta^{+\mu}{\partial^-}_\mu\phi^*(x)
 +\frac{i}{4}\theta^-\theta^{-\mu}{\partial^+}_\mu\phi^*(x)
 -\frac{1}{2}\theta^+\theta^{-\mu}{K^+}_\mu(x)\\
& \quad +\frac{i}{8}\theta^+\theta^{+\rho}{\theta^-}_\rho{\partial^{-\mu}}
{\xi^+}_\mu(x)
 +\frac{i}{4}\theta^+\theta^{-\mu}\theta^-{\partial^+}_\mu\xi^*(x)\\
& \quad -\frac{1}{8}\theta^+\theta^-\theta^{+\mu}{\theta^-}_\mu
\partial^2\phi^*(x)\\
&=\phi^*(y^+)
 -\frac{1}{2}\theta^{-\mu}{\xi^+}_\mu(y^+)
 +\theta^+ \left(
  -\xi^*(y^+)
  -\frac{1}{2}\theta^{-\mu}{K^+}_\mu(y^+)
 \right) \\
 &=\Phi^+ (y^{+\mu},\theta^{-\mu},\theta^+)  = 
 e^{\theta^+ Q^- + \frac{1}{2}\theta^{-\mu} Q_{\mu}^+}\phi(y^+),
\end{split}
\label{eq:Phi+}
\end{equation}
where ${\xi^+}_\mu=({\xi^-}_\mu)^*$, ${K^+}_\mu=({K^-}_\mu)^*$, and
$y^{+\mu}=x^\mu-\frac{i}{2}\theta^+\theta^{+\mu}+\frac{i}{2}\theta^-\theta^{
-\mu}$. Indeed $\Phi^+$ satisfy the following conditions:
\begin{align}
D^+\Phi^+ \ \ = \ \  {D^-}_\mu\Phi^+=0, 
\end{align}
since 
\begin{eqnarray}
D^+y^{+\mu} &=& D^+\theta^+ = D^+\theta^{-\mu} = 0 \nonumber \\ 
D^-_\mu y^{+\nu} &=& D^-_\mu\theta^+ = D^-_\mu\theta^{-\nu} = 0.
\end{eqnarray}
Similarly we can find N=2 supersymmetry transformation laws of the 
component fields
$\phi^*$, ${\xi^+}_\mu$, $\xi^*$, and ${K^+}_\mu$(see Table 7).

\begin{table}[htbp]
\begin{center}
\begin{tabular}{c|c|c|c|c}
\hline
$\phi^A$ & $Q^-\phi^A$ & ${Q^-}_\mu\phi^A$ & $Q^+\phi^A$ 
& ${Q^+}_\mu\phi^A$\\
\hline
$\phi$ & $0$ & ${\xi^-}_\mu$ & $\xi$ & $0$ \\
${\xi^-}_\rho$ & $-i{\partial^-}_\rho \phi$ & $0$
 & $-{K^-}_\rho$ & $0$ \\
$\xi $ & $0$ & ${K^-}_\mu$
 & $0$ & $-i{\partial^+}_\mu \phi$ \\
${K^-}_\rho$ & $-i{\partial^-}_\rho \xi$ & $0$
  & $0$ & $i{\partial^+}_\mu {\xi^-}_\rho$ \\
\hline
$\phi^*$ & $-\xi^*$ & $0$ & $0$ & $-{\xi^+}_\mu$ \\
${\xi^+}_\rho$ & $-{K^+}_\rho$ & $0$
  & $i{\partial^+}_\rho \phi^*$ & $0$ \\
$\xi^* $ & $0$ & $i{\partial^-}_\mu \phi^*$
  & $0$ & ${K^+}_\mu$ \\
${K^+}_\rho$ & $0$ & $i{\partial^-}_\mu {\xi^+}_\rho$
  & $-i{\partial^+}_\rho \xi^*$ & $0$ \\
\hline
\end{tabular}
\caption{N=2 super-transformation of chiral twisted super multiplet. 
$(\phi,{\xi^-}_\rho, \xi, {K^-}_\rho)$ and 
$(\phi^*$, ${\xi^+}_\rho$, $\xi^*$, and ${K^+}_\rho)$.}
\end{center}
\end{table}

Since $\theta^4$ component of $\Phi^+\Phi$ transforms into a space
derivative, the following action is invariant under N=2 
super-transformation:
\begin{align}
S&=\int d^2x \int d^4\theta\ \Phi^+ \Phi \nonumber\\
 &=\int d^2x \biggl[
  \frac{1}{4}\partial^{+\mu}\phi^* \partial_{-\mu}\phi
  -\frac{i}{4}\partial^{-\mu}{\xi^+}_\mu \xi
  -\frac{i}{4}\partial^{+\mu}{\xi^-}_\mu \xi^*
  -\frac{1}{4}{K^-}_\mu K^{+\mu}
 \biggr]\nonumber\\
&=\int d^2x \biggl[
 \frac{1}{4}\partial_\mu\phi_i\partial^\mu\phi^i
 -\frac{i}{2}\partial^\mu\xi_\mu \psi
 -\frac{i}{2}\epsilon^{\mu\nu}\partial_\mu\xi_\nu \tilde{\psi}
 -\frac{1}{4}K_\mu K^\mu
\biggr],
\end{align}
where in the last equality we use the following redefinitions of fields:
\begin{equation}
\begin{split}
\phi &= \frac{1}{\sqrt{2}}(\phi_1-i\phi_2),
 K_\mu=\frac{1}{\sqrt{2}}({K^+}_\mu+{K^-}_\mu),\\
\xi&=\frac{1}{\sqrt{2}}(\psi-i\tilde{\psi}),\
 \xi_\mu=\frac{1}{\sqrt{2}}({\xi^+}_\mu+{\xi^-}_\mu).
\end{split}
\label{eq:WZ action 3}
\end{equation}
Here again, the set of fields $(\psi,\xi_\mu,\tilde{\psi})$ in 
(\ref{eq:WZ action 3}) lead matter fermion via Dirac-K\"ahler fermion 
mechanism:
\begin{align}
-\int d^2x\ \left(
  \frac{i}{2}\partial^\mu\xi_\mu \psi
  +\frac{i}{2}\epsilon^{\mu\nu}\partial_\mu\xi_\nu \tilde{\psi}
 \right)
=\int d^2x\
  \text{Tr} \left(
   \frac{i}{2}\overline{\xi} \gamma^\mu \partial_\mu \xi
  \right),
\end{align}
where
\begin{align}
\xi_{\alpha\beta}
=\frac{1}{2}\left(
 \mathbf{1}\psi+\gamma^\mu\xi_\mu+\gamma^5\tilde{\psi}
\right)_{\alpha\beta},
\label{DKtr3}
\end{align}
and $\overline{\xi}=C\xi^TC=\xi^T$.

As we will see later, these types of chiral superfields will appear when the
Yang-Mills type action will be introduced. And then the fermionic contents
$(\xi,\xi^*,{\xi^+}_\mu,{\xi^-}_\mu)$ will lead to matter fermion via 
Dirac-K\"ahler fermion mechanism again.

\subsection{Vector superfields and gauge transformation}
Vector superfield is defined as the superfield satisfying the condition
\begin{align}
V=V^*,
\label{vector_ss_condition}
\end{align}
and can be expanded as follows:
\begin{align}
V&(x^\mu,\theta^+,\theta^{+\mu},\theta^-,\theta^{-\mu})\nonumber\\
&=
 L(x)
 +\frac{1}{2}\theta^{+\mu}{\eta^-}_\mu(x)
 -\frac{1}{2}\theta^{-\mu}{\eta^+}_\mu(x)
 +\theta^-\eta(x)
 -\theta^+\eta^*(x)\nonumber\\
& \quad
 +\frac{1}{2}\theta^-\theta^{+\mu}{M^-}_\mu(x)
 -\frac{1}{2}\theta^+\theta^{-\mu}{M^+}_\mu(x)\nonumber\\
& \quad
 +\frac{1}{2}\theta^+\theta^{+\mu}{\omega^-}_\mu(x)
 -\frac{1}{2}\theta^-\theta^{-\mu}{\omega^+}_\mu(x)
 +\theta^+\theta^-A(x)
 +\frac{1}{2}\theta^{+\mu}{\theta^-}_\mu B(x)\nonumber\\
& \quad
 +\frac{1}{2}\theta^+\theta^{+\mu}\theta^-
  \left({\lambda^-}_\mu(x)+\frac{i}{2}{\partial^-}_\mu\eta(x)\right)
 -\frac{1}{2}\theta^+\theta^{-\mu}\theta^-
\left({\lambda^+}_\mu(x)-\frac{i}{2}
{\partial^+}_\mu\eta^*(x)\right)\nonumber\\
& \quad +\frac{1}{2}\theta^-\theta^{+\rho}{\theta^-}_\rho
  \left(\lambda(x)+\frac{i}{4}{\partial^+}_\mu\eta^{-\mu}(x)\right)
 -\frac{1}{2}\theta^+\theta^{+\rho}{\theta^-}_\rho
\left(\lambda^*(x)-\frac{i}{4}{\partial^-}_\mu\eta^{+\mu}(x)\right)\nonumber
\\
& \quad
 +\frac{1}{2}\theta^+\theta^-\theta^{+\rho}{\theta^-}_\rho
  \left(D(x)-\frac{1}{4}\partial^2L(x)\right),
\end{align}
where $L$, $A$, $B$, and $D$ must all be real.

We consider the following transformation:
\begin{align}
V \rightarrow V^\prime=V+\Lambda+\Lambda^+,
\label{eq:V trans}
\end{align}
where $\Lambda^+ = \Lambda^*$ and thus $\Lambda + \Lambda^+$ satisfies the 
vector superspace condition (\ref{vector_ss_condition}).
Here $\Lambda$ and $\Lambda^+$ are chiral superfields,
\begin{equation}
\begin{split}
\Lambda&=v(x)
 +\frac{1}{2}\theta^{+\mu}{\zeta^-}_\mu(x)
 +\theta^-\zeta(x)\\
& \quad
 +\frac{i}{4}\theta^+\theta^{+\mu}{\partial^-}_\mu v(x)
 -\frac{i}{4}\theta^-\theta^{-\mu}{\partial^+}_\mu v(x)
 +\frac{1}{2}\theta^-\theta^{+\mu}{N^-}_\mu(x)\\
& \quad +\frac{i}{8}\theta^-\theta^{+\rho}{\theta^-}_\rho
{\partial^{+\mu}}{\zeta^-}_\mu(x)
 +\frac{i}{4}\theta^+\theta^{+\mu}\theta^-{\partial^-}_\mu\zeta(x)\\
& \quad
 -\frac{1}{8}\theta^+\theta^-\theta^{+\mu}{\theta^-}_\mu \partial^2 v(x),
\end{split}
\label{eq:Lambda}
\end{equation}
and
\begin{equation}
\begin{split}
\Lambda^+
&= \Lambda^* = v^*(x)
 -\frac{1}{2}\theta^{-\mu}{\zeta^+}_\mu(x)
 -\theta^+\zeta^*(x)\\
& \quad -\frac{i}{4}\theta^+\theta^{+\mu}{\partial^-}_\mu v^*(x)
 +\frac{i}{4}\theta^-\theta^{-\mu}{\partial^+}_\mu v^*(x)
 -\frac{1}{2}\theta^-\theta^{+\mu}{N^+}_\mu(x)\\
& \quad +\frac{i}{8}\theta^+\theta^{+\rho}{\theta^-}_\rho{\partial^{-\mu}}
{\zeta^+}_\mu(x)
 +\frac{i}{4}\theta^+\theta^{-\mu}\theta^-{\partial^+}_\mu\zeta^*(x)\\
& \quad -\frac{1}{8}\theta^+\theta^-\theta^{+\mu}{\theta^-}_\mu 
\partial^2v^*(x),
\end{split}
\label{eq:Lambda+}
\end{equation}
where $D^+\Lambda={D^-}_\mu\Lambda=0$ and
$D^-\Lambda^+={D^+}_\mu\Lambda^+=0$. Under this transformation, the
component fields of $V$ transform as follows:
\begin{equation}
\begin{split}
L & \rightarrow L+ v+v^*,\\
{\eta^\pm}_\mu & \rightarrow {\eta^\pm}_\mu+{\zeta^\pm}_\mu,\\
\eta & \rightarrow \eta+\zeta,\\
{M^\pm}_\mu & \rightarrow {M^\pm}_\mu+{N^\pm}_\mu, \\
{\omega^\pm}_\mu & \rightarrow {\omega^\pm}_\mu
 +\frac{i}{2}{\partial^\pm}_\mu(v-v^*) ,\\
A & \rightarrow A ,\\
B & \rightarrow B ,\\
{\lambda^\pm}_\mu & \rightarrow {\lambda^\pm}_\mu, \\
\lambda & \rightarrow \lambda, \\
D & \rightarrow D.
\end{split}
\label{eq:V trans 2}
\end{equation}
While $\omega_\mu=\frac{1}{\sqrt{2}}({\omega^+}_\mu+{\omega^-}_\mu)$ 
transforms as:
\begin{align}
\omega_\mu \rightarrow \omega_\mu+\partial_\mu \text{Im}(v),
\end{align}
where we can identify $\omega_\mu$ and Im$(v)$ as the gauge field and the
gauge parameter, respectively. Then the transformation (\ref{eq:V trans}),
or (\ref{eq:V trans 2}), can be regarded as the supersymmetric
generalization of the gauge transformation.

We introduce the following superfields:
\begin{align}
W &=\frac{1}{2}D^-{D^-}_\mu D^{+\mu} V,\\
W_\mu &={D^+}_\mu D^-D^+ V.
\end{align}
These are chiral superfields:
\begin{equation}
\begin{split}
D^-W &=0,\ {D^+}_\mu W =0,\\
D^-W_\mu &=0,\ {D^+}_\rho W_\mu =0,
\end{split}
\end{equation}
and they are gauge invariant under the gauge transformation (\ref{eq:V
trans}),
\begin{align}
W &\rightarrow \frac{1}{2}D^+{D^+}_\mu D^{-\mu} (V+\Lambda+\Lambda^+)=W,\\
W_\mu &\rightarrow {D^-}_\mu D^+D^- (V+\Lambda+\Lambda^+)=W_\mu.
\end{align}

In the following subsection, we introduce the super and gauge invariant action
described by the chiral super fields $(\Phi,W,W_\mu)$, anti-chiral 
super field $\Phi^+$ and the vector superfield $V$.

\newpage
\subsection{Twisted super Yang-Mills action}
We introduce the following action:
\begin{align}
S=\frac{1}{2}\int d^2x \int d\theta^{+\mu} d\theta^- W W_\mu
+4\int d^2x \int d^4\theta\ \Phi^+ e^{gV} \Phi,
\label{eq:SYM action 2}
\end{align}
where $g$ is a constant, $\Phi$ and $\Phi^+$ are chiral superfields
(\ref{eq:Phi}) and (\ref{eq:Phi+}), respectively. The gauge transformations
of the chiral superfield are defined as follows:
\begin{equation}
\begin{split}
\Phi &\rightarrow \Phi^\prime=e^{-g\Lambda}\Phi,\\
\Phi^+ &\rightarrow \Phi^{\prime +}=\Phi^+e^{-g\Lambda^+}.\\
\end{split}
\label{eq:trans chiral}
\end{equation}
Since $W$ and $W_\mu$ are chiral superfields, $\theta^{+\mu}\theta^-$
component of $WW_\mu$ transforms into a space derivative, and therefore the
first term in (\ref{eq:SYM action 2}) is super and gauge invariant. The
second term is trivially super and gauge invariant.

By using the degrees of freedom of the gauge parameters except Im$(v)$, we
can take a special gauge, Wess-Zumino gauge, in which $L$, ${\eta^\pm}_\mu$,
$\eta$, and ${M^\pm}_\mu$ are all set to be zero by adjusting the parameters; 
$v+v^*, \zeta^{\pm}{}_\mu, \zeta$, and $N^{\pm}{}_\mu$ respectively:
\begin{align}
V&(x^\mu,\theta^+,\theta^{+\mu},\theta^-,\theta^{-\mu})\nonumber\\
&=
 \frac{1}{2}\theta^+\theta^{+\mu}{\omega^-}_\mu
 -\frac{1}{2}\theta^-\theta^{-\mu}{\omega^+}_\mu
 +\theta^+\theta^-A
 +\frac{1}{2}\theta^{+\mu}{\theta^-}_\mu B\nonumber\\
& \quad
 +\frac{1}{2}\theta^+\theta^{+\mu}\theta^-{\lambda^-}_\mu
 -\frac{1}{2}\theta^+\theta^{-\mu}\theta^-{\lambda^+}_\mu\nonumber\\
& \quad
 +\frac{1}{2}\theta^-\theta^{+\rho}{\theta^-}_\rho\lambda
 -\frac{1}{2}\theta^+\theta^{+\rho}{\theta^-}_\rho\lambda^*\nonumber\\
& \quad
 +\frac{1}{2}\theta^+\theta^-\theta^{+\rho}{\theta^-}_\rho D.
\label{eq:V}
\end{align}
Then $W$ and $W_\mu$ turn out as follows:
\begin{align}
W(y^\mu,\theta^{+\mu},\theta^-) &= \lambda^*
 -\theta^-\left(
  D
  -\frac{i}{4}\partial^{+\rho}{\omega^-}_\rho
  +\frac{i}{4}\partial^{-\rho}{\omega^+}_\rho \right)
 \nonumber\\
& \qquad
 +\frac{i}{2}\theta^{+\mu}{\partial^-}_\mu B
-\frac{i}{2}\theta^-\theta^{+\mu}{\partial^-}_\mu \lambda,
\label{eq:W}\\
W_\mu(y^\mu,\theta^{+\mu},\theta^-) &= -{\lambda^+}_\mu+
i\theta^-{\partial^+}_\mu A
\nonumber\\
& \qquad
 +{\theta^+}_\mu \left(
  D
  +\frac{i}{4}\partial^{+\rho}{\omega^-}_\rho
  -\frac{i}{4}\partial^{-\rho}{\omega^+}_\rho
 \right)
+\frac{i}{2}{\theta^+}_\mu\theta^-\partial^{+\rho}{\lambda^-}_\rho.
\label{eq:W_mu}
\end{align}

Once we take the Wess-Zumino gauge, the structure of supersymmetry breaks down.
The action is, however, invariant under a new supersymmetry transformation 
combined with the 
gauge transformation; $\delta=\delta_{super}+\delta_{gauge}$. We can
redefine this combined transformation as the new super transformation, where
the new supersymmetry algebra closes up to the gauge transformation.

Now the action (\ref{eq:SYM action 2}) in the Wess-Zumino gauge turns out as:
\begin{align}
S=\int d^2x \biggl[
& 2\left(
\frac{i}{4}\partial^{+\mu}{\omega^-}_\mu-\frac{i}{4}\partial^{-\mu}
{\omega^+}_\mu\right)^2
+i\partial^{-\mu}{\lambda^+}_\mu \lambda
+i\partial^{+\mu}{\lambda^-}_\mu \lambda^*
+{\partial^+}_\mu A \partial^{-\mu}B - 2D^2
 \nonumber\\
&  -2\partial^2\phi^* \phi
  -{K^-}_\mu K^{+\mu}
  -i\partial^{-\mu}{\xi^+}_\mu \xi
  -i\partial^{+\mu}{\xi^-}_\mu \xi^*
  \nonumber\\
& +g\biggl(
  -\frac{i}{2}\phi{\omega^-}_\mu\partial^{+\mu}\phi^*
  +\frac{i}{2}\phi^*{\omega^-}_\mu\partial^{+\mu}\phi
  -\frac{i}{2}\phi{\omega^+}_\mu\partial^{-\mu}\phi^*
  +\frac{i}{2}\phi^*{\omega^+}_\mu\partial^{-\mu}\phi
  \nonumber\\
 & \qquad \qquad
  -\xi{\omega^-}_\mu\xi^{+\mu}
  +\xi^*{\omega^+}_\mu\xi^{-\mu}
  +A{\xi^-}_\mu\xi^{+\mu}
  -2B\xi\xi^*
  \nonumber\\
 & \qquad \qquad
  +\lambda^{+\mu}{\xi^-}_\mu\phi^*
  -\lambda^{-\mu}{\xi^+}_\mu\phi
  +2\lambda^*\xi\phi^*
  -2\lambda\xi^*\phi
  +2D\phi\phi^*
 \biggr)
 \nonumber\\
& +g^2 \left(
  {\omega^-}_\mu\omega^{+\mu}\phi\phi^*
  +2AB\phi\phi^*
 \right)
 \biggr].
\end{align}

We redefine the component fields as:
\begin{equation}
\begin{split}
\phi&=\frac{1}{\sqrt{2}}(\phi_1-i\phi_2),
  \omega_\mu =\frac{1}{\sqrt{2}}({\omega^+}_\mu+{\omega^-}_\mu),\\
\lambda&=\frac{1}{\sqrt{2}}(\rho-i\tilde{\rho}),
  \lambda_\mu =\frac{1}{\sqrt{2}}({\lambda^+}_\mu+{\lambda^-}_\mu),\\
\xi&=\frac{1}{\sqrt{2}}(C_1-iC_2),
  K_\mu=\frac{1}{\sqrt{2}}({K^+}_\mu+{K^-}_\mu),\\
\chi_{\mu 1} &=\frac{1}{\sqrt{2}}({\xi^+}_\mu+{\xi^-}_\mu),
  \chi_{\mu 2}=\frac{1}{\sqrt{2}}\epsilon_{\mu\nu}(\xi^{+\nu}+\xi^{-\nu}),
\end{split}
\end{equation}
where $\chi_{\mu i}$ is the anti-selfdual field which obeys the condition
$\chi_{\mu i}=-\epsilon_{\mu\nu}\epsilon_{ij}\chi^{\nu j}$.

Then the action can be rewritten as follows:
\begin{equation}
\begin{split}
S=\int d^2x \biggl[
&\frac{1}{2}F^{\mu\nu}F_{\mu\nu}
+2i\partial^\mu\lambda_\mu \rho
+2i\epsilon^{\mu\nu}\partial_\mu\lambda_\nu \tilde{\rho}
-2A\partial^2B\\
&-2D^2
+gD\phi_i\phi^i
-K_\mu K^\mu\\
& +(D_\mu\phi)_i(D^\mu\phi)^i
+2i\chi_{\mu i}(D_\mu C)^i\\
&+\sqrt{2}ig\lambda^\mu\epsilon^{ij}\chi_{\mu i}\phi_j
+\sqrt{2}ig(
\rho\epsilon^{ij}C_i\phi_j
+\tilde{\rho}C_i\phi^i)\\
&+\frac{i}{2}gA\epsilon^{ij}\chi_{\mu i}{\chi^\mu}_j
-igB\epsilon^{ij}C_iC_j
+g^2AB\phi_i\phi^i
\biggr],
\end{split}
\label{eq:off-shell KT action}
\end{equation}
where $F_{\mu\nu}=\partial_\mu\omega_\nu-\partial_\nu\omega_\mu$, and
$(D_\mu\phi)_i=\partial_\mu\phi_i-\frac{1}{\sqrt{2}}g\omega_\mu\epsilon_{ij}
\phi^j$. This action is the off-shell N=2 supersymmetry invariant extended 
version of the action
(\ref{eq:KT action}) obtained from partially gauge fixing of the generalized
topological Yang-Mills action with the instanton conditions. Indeed,
integrating $D$ and $K_\mu$, the action (\ref{eq:off-shell KT action}) turns
out to be equivalent to the action (\ref{eq:KT action}):
\begin{equation}
\begin{split}
S=\int d^2x \biggl[
&\frac{1}{2}F^{\mu\nu}F_{\mu\nu}
+2i\partial^\mu\lambda_\mu \rho
+2i\epsilon^{\mu\nu}\partial_\mu\lambda_\nu \tilde{\rho}
-2A\partial^2B\\
&+|\phi|^4
 +(D_\mu\phi)_i(D^\mu\phi)^i
+2i\chi_{\mu i}(D_\mu C)^i\\
&+4i\lambda^\mu\epsilon^{ij}\chi_{\mu i}\phi_j
+4i(
\rho\epsilon^{ij}C_i\phi_j
+\tilde{\rho}C_i\phi^i)\\
&+\sqrt{2}iA\epsilon^{ij}\chi_{\mu i}{\chi^\mu}_j
-2\sqrt{2}iB\epsilon^{ij}C_iC_j
+8AB|\phi|^2
\biggr],
\end{split}
\end{equation}
where we set $g=2\sqrt{2}$. Here the sets $(\phi_i,C_i,\chi_{\mu i})$ and
$(A,B,\omega_\mu,\rho,\lambda_\mu,\tilde{\rho})$ are the N=2 chiral
multiplet and the N=2 vector multiplet, respectively, and their gauge
transformations are given by the transformations (\ref{eq:trans chiral}) and
(\ref{eq:V trans}) in the Wess-Zumino gauge:
\begin{equation}
\begin{split}
\delta_{gauge}(\phi_i,C_i,\chi_{\mu i})
 &=2\text{Im}(v) \epsilon_{ij}(\phi^j,C^j,{\chi_\mu}^j),\\
\delta_{gauge}\omega_\mu&=\partial_\mu v,\\
\delta_{gauge}(A,B,\rho,\lambda_\mu,\tilde{\rho})&=0,
\end{split}
\label{eq:SO(2) trans 2}
\end{equation}
where Im$(v)$ is the gauge parameter. These transformation of the fields 
is  the $SO(2)$ gauge transformation (\ref{eq:SO(2) trans}).


\section{Twisting as Dirac-K\"ahler fermion mechanism}
\setcounter{equation}{0}

We have observed in the quantization of the generalized topological 
Yang-Mills theory that a linear 
combination of the ghosts of shift symmetry of the topoligical nature 
of generalized Yang-Mills; $\tilde{C}_\mu$ and $\tilde{C}_a$, the 
anti-ghosts of instanton gauge fixing; $\chi_{\mu a}$ and $\lambda$, and 
the Lagrange multiplier of ghost symmetry; $\rho$, turn into matter 
fermions via the twisting procedure which we identify as 
Dirac-K\"ahler fermion mechanism\cite{KT}. 
In other words the vector or tensor suffices of ghost fields 
and Lagrange multiplier can be transformed into spinor suffices as 
in the eqs. (\ref{DKtr01}) and (\ref{DKtr02}): 
\begin{align}
\psi_{\alpha i} &= \tfrac{1}{2}(\rho+
\gamma^\mu\tilde{C}_\mu-\gamma^5\lambda)_{\alpha i}, 
\label{DKtr5_1}\\
\chi_{\alpha i} &= \tfrac{1}{2}(-\tilde{C}_{1}+\gamma^\mu\chi_{\mu
1}-\gamma^5\tilde{C}_{2})_{\alpha i}. \label{DKtr5_2}
\end{align}
It works exactly in the same way for the fields of (\ref{DKtr1}) and 
(\ref{DKtr2}).
\begin{eqnarray}
\xi_{\alpha i}
&=&\frac{1}{2}\left(
 \mathbf{1}\chi+\gamma^\mu\phi_\mu+\gamma^5\tilde{\chi}
\right)_{\alpha i},  \\
\psi_{\alpha i}
&=&\frac{1}{2}\left(
 \mathbf{1}\psi+\gamma^\mu\psi_\mu+\gamma^5\tilde{\psi}
\right)_{\alpha i}.
\end{eqnarray}

The super charges of N=2 supersymmetry and the twisted N=2 super charges 
have the following relation:
\begin{align}
Q_{\alpha i}= \left(\mathbf{1} Q + \gamma^\mu Q_\mu + \gamma^5
\tilde{Q}\right)_{\alpha i}.
\end{align}
This leads to the following algebra including the angular momentum generator 
$J$ and R-symmetry generator $R$:
\begin{equation}
\begin{split}
[J,Q_{\alpha i}]&=\tfrac{i}{2}(\gamma^5)_{\alpha}{}^{\beta}Q_{\beta i},\\
[R,Q_{\alpha i}]&=\tfrac{i}{2}(\gamma^5)_i{^j}Q_{\alpha j}
\end{split}
\label{eq:N=2 algebra2}
\end{equation}
and the algebra including the angular momentum generator of the twisted 
space $J^\prime$ and $R$:
\begin{equation}
\begin{split}
[J^\prime,Q]&=[J^\prime,\tilde{Q}]=0,\
[J^\prime,Q_\mu]=i\epsilon_{\mu\nu}Q^\nu,\\
[R,Q]&=\tfrac{i}{2}\tilde{Q},\
[R,Q_\mu]=\tfrac{i}{2}\epsilon_{\mu\nu}Q^\nu,\
[R,\tilde{Q}]=-\tfrac{i}{2}Q.
\end{split}
\label{eq:N=2 T-algebra2}
\end{equation}
Those generators have the following relation:
\begin{equation}
J^\prime = J + R.
\end{equation}

As we can see from the above relations, the angular momentum generator $J$ 
generates the Lorentz rotation of the spinor suffix and thus the spinor 
fields $\psi_{\alpha i},\chi_{\alpha i}$ and $\xi_{\alpha i},\psi_{\alpha i}$ 
have half integer spin while the R-symmetry generator $R$ rotates the "flavor" 
suffix $i$ of those fermion fields.   
It is, however, interesting to recognize that the twisted angular momentum 
generator $J^\prime$ rotates the scalar, vector and tensor suffix of 
twisted super charges.  
In other words the fermion fields with scalar, vector and tensor 
suffixes transform as integer spin fermion fields. 
Thus the R-symmetry generator $R$ plays the role of 
shifting the integer spin of ghost fermions into the half integer spin of 
the matter fermions. 

This situation can be more clearly seen by the following concrete example. 
The Lorentz transformation on the Dirac-K\"ahler field $\psi$ of 
(\ref{DKtr5_1}) induced by $J'$ is given by 
\begin{eqnarray}
 \delta_{J'}\psi &=& \frac{1}{2}
                     \Big(i{\epsilon_\mu}^\nu\tilde{C}_\nu\gamma^\mu\Big) 
                     \nonumber \\
                 &=& i\frac{1}{2}{[\gamma_5, \psi]}.
\end{eqnarray}
On the other hand the Lorentz transformation induced by $J=J'-R$ is 
\begin{eqnarray}
 \delta_J\psi &=& \delta_{J'}\psi - \delta_R\psi \nonumber \\
              &=& i\frac{1}{2}\gamma_5\psi, 
\end{eqnarray}
which precisely coincides with the Lorentz transformation of spinor field 
since the R-symmetry rotation $\delta_R\psi$ is the rotation of the 
"flavor" suffix $i$ of the matter fermions. 
This shows that Dirac-K\"ahler fermion mechanism is essentially related 
to the twisting procedure with $N=2$ supersymmetry.


\setcounter{equation}{0}
\renewcommand{\theequation}{\arabic {section}.\arabic{equation}}

\section{Conclusions and Discussions}
We have explicitly shown that the BRST charge; $s$, the vector supersymmetry 
charge; $s_\mu$, and pseudo scalar supersymmetry charge; 
$\tilde{s}$, constitute the twisted super 
charges of D=N=2 super symmetry and are related to the N=2 supersymmetry 
charges; $Q$, in the following simple relation:
\begin{align}
Q= \left(\mathbf{1} s + \gamma^\mu s_\mu + \gamma^5 \tilde{s}\right), 
\label{supercharge-twist2}
\end{align}
which has the same algebraic structure as the Dirac-K\"ahler fermion 
formulation. R-symmetry of N=2 super algebra is the "flavor" 
symmetry of the Dirac-K\"ahler field. 

Based on this supersymmetry algebra we have proposed the 
twisted D=N=2 superspace formalism. We have defined the twisted 
superspace in terms of explicitly defined differential operators. 
We first have constructed models with the chiral superfield set 
$(\Psi,\overline{\Psi})$:
\begin{align}
S&=(-i)^{\epsilon_\Psi}\int d^2x d^4\theta \ 
\hbox{Tr} \ (\overline{\Psi}\Psi),
\end{align}
where $\Psi$ and $\overline{\Psi}$ are chiral and anti-chiral superfields 
for Abelian case 
while the chiral condition is broken for the non-Abelian case. 
We have explicitly shown that a fermionic bilinear form of twisted N=2 chiral 
and anti-chiral superfields 
is equivalent to the quantized version of BF theory with the Landau 
type gauge fixing while a bosonic bilinear form leads to the 
N=2 Wess-Zumino action after nontrivial change of field variables. 
In showing these equivalences we have found that the
ghost fields turn into matter fermion by the twisting mechanism. We claim that 
this twisting mechanism is nothing but the Dirac-K\"ahler fermion
formulation, which is the essence of the twisted superspace formulation.

Secondly, we have constructed the super and gauge invariant action with the
chiral superfields $(\Phi,\Phi^+)$ 
and the vector superfield $V$,
\begin{align}
S=\int d^2x \int d\theta^{+\mu} d\theta^- W W_\mu
+4\int d^2x \int d^4\theta\ \Phi^+ e^{gV} \Phi.
\end{align}
We have shown that the action in the Wess-Zumino gauge turns out to be 
off-shell twisted N=2 supersymmetric action which leads to the partially 
gauge fixed action with the instanton conditions of the generalized 
topological Yang-Mills theory after integrating auxiliary fields. 
The close relations between the twisting mechanism and Dirac-K\"ahler fermion 
mechanism can be more explicitly seen in this example. 
The role of the R-symmetry and the change of the spin structre from 
ghost related fermions to matter fermions in the Dirac-K\"ahler fermion 
mechanism are explicitly shown.  

The Dirac-K\"ahler fermion is formulated by introducing all 
the degrees of inhomogeneous differential forms which are then 
transformed into Dirac-K\"ahler fermion fields possessing the spinor 
and "flavor" suffixes. 
It is interesting to recognize that the generalized gauge theory 
possesses all the degrees of differential forms as gauge parameters which 
then turn into ghost fields after the quantization of the generalized 
gauge fields\cite{KOS,KSTU}. 
We expect that the quantized generalized gauge theory may lead to 
extended supersymmetry invariant actions even in other dimensions than two 
dimensions.

As we can see from the formulation, the generalization of the twisted 
superspace formalism into four dimensions is very important proplem to 
pursue since we expect to obtain off-shell N=4 supersymmetric actions 
in four dimensions. 
There are already some investigations on D=N=4 formulation 
of the topological field theories\cite{Yam,VW,LabLoz,LSSTV}.
Along the line of the formulation of this paper we can indeed generalize 
it into four dimensions, however, it is not easy to find the known 
actions having N=4 supersymmetry\cite{KKM}.

Another important issue to recognize is that Dirac-K\"ahler fermion is 
a curved spacetime version of Kogut-Susskind fermion~\cite{KSuss,Suss} 
or staggered fermion~\cite{KS} and thus a natural framework 
of the lattice fermion formulation. 
It was shown in the recent paper\cite{KK} that the Dirack-K\"ahler 
fermion formulation can be defined in 
unambiguous way in terms 
of Clifford product by introducing noncommutative 
differential form on the lattice. 
Here the notorious problem of the difficulty of the Leipnitz rule 
on the lattice is avoided by paying the price of introducing 
noncommutativity. 
We would like to show that the supersymmetry on the lattice with 
the noncommutativity will be realized by using the formulation of 
the twisted superspace formalism of this article.

\vspace{1cm}

\textbf{\Large Acknowledgements}

We would like to thank T. Tsukioka and Y. Watabiki for useful discussions 
at the early stage of this work. 
This work is supported in part by Japanese 
Ministry of Education, Science, Sports and Culture under the grant number 
13640250 and 13135201.

\begin{center}
{\Large \textbf{Appendix}\\[0pt]}
\end{center}

\noindent
{\Large \textbf{A. Notations }\\[0pt]}

\renewcommand{\theequation}{A.\arabic{equation}}
\setcounter{equation}{0}

Throughout this paper, we consider the 
two-dimensional Euclidean space-time, where the $\gamma$- matrices satisfy
\begin{equation}
\begin{split}
\{\gamma^\mu,\gamma^\nu\}&=2\delta^{\mu\nu}\mathbf{1},\\
\{\gamma^5,\gamma^\mu\}&=0,\\
C\gamma^\mu C^{-1}&=\gamma^{\mu T},
\end{split}
\end{equation}
where
$\gamma^5=\frac{1}{2}\epsilon_{\mu\nu}\gamma^\mu\gamma^\nu=\gamma^1\gamma^2$
, $\epsilon_{12}=-\epsilon_{21}=\epsilon^{12}=-\epsilon^{21}=1$, and $C$ is
charge conjugation matrix which can be taken as $C=\mathbf{1}$ in the
representation $\gamma^\mu=(\sigma^3,\sigma^1)$.


\end{document}